\documentclass[10pt,superscriptaddress,eqsecnum,twocolumn]{revtex4}
\usepackage{graphicx} 
\usepackage{color} 
\usepackage{amsthm}
\usepackage{amsmath}
\usepackage{amssymb}
\usepackage{bbm,amsfonts}
\usepackage{setspace}
\usepackage{nomencl}
\usepackage{amstext} 

\newcommand{\Eop}{\mathcal{E}}
\newcommand{\Rop}{\mathcal{R}}

\newcommand{\Id}{\mathbbm{1}}

\newcommand{\braa}{\langle \psi |}
\newcommand{\kett}{|\psi\rangle}

\newcommand{\Hilb}{\mathcal{H}}

\newcommand{\Clif}{\text{Clif}_n}

\newcommand{\tr}{{\mathrm{tr}}}

\newcommand{\pure}{\mathbb{CP}^{d-1}}
\newcommand{\Fg}{\mathcal{F}_g(m,\psi)}
\newcommand{\Fgzero}{\mathcal{F}_g^{(0)}(m,|\psi\rangle)}
\newcommand{\Fgone}{\mathcal{F}_g^{(1)}(m,|\psi\rangle)}
\newcommand{\Fgtwo}{\mathcal{F}_g^{(2)}(m,|\psi\rangle)}
\newcommand{\Fgk}{\mathcal{F}_g^{(k)}(m,\psi)}
\newcommand{\Fgkpone}{\mathcal{F}_g^{(k+1)}(m,\psi)}

\newcommand{\Fgvec}{\mathcal{F}_g^{\vec{i_m}}(m,|\psi\rangle)}

\newtheorem{definition}{Definition}

\def\>{\rangle}
\def\<{\langle}

\begin{document}

\title{Characterizing Quantum Gates via Randomized Benchmarking}

\author{Easwar Magesan}
\affiliation{Department of Applied Mathematics, University of Waterloo, Waterloo, ON N2L 3G1, Canada}	
\affiliation{Institute for Quantum Computing, University of Waterloo, Waterloo, ON N2L 3G1, Canada}
\author{Jay M. Gambetta}
\affiliation{IBM T.J. Watson Research Center, Yorktown Heights, NY 10598, USA}
\author{Joseph Emerson}
\affiliation{Department of Applied Mathematics, University of Waterloo, Waterloo, ON N2L 3G1, Canada}	
\affiliation{Institute for Quantum Computing, University of Waterloo, Waterloo, ON N2L 3G1, Canada}

\begin{abstract}
We describe and expand upon the scalable randomized benchmarking protocol proposed in \emph{Phys. Rev. Lett. 106, 180504 (2011)} which provides a method for benchmarking quantum gates and estimating the gate-dependence of the noise. The protocol allows the noise to have weak time and gate-dependence, and we provide a sufficient condition for the applicability of the protocol in terms of the average variation of the noise. We discuss how state preparation and measurement errors are taken into account and provide a complete proof of the scalability of the protocol. We establish a connection in special cases between the error rate provided by this protocol and the error strength measured using the diamond norm distance.
\end{abstract}

\maketitle


\section{Introduction}

Quantum computers promise an exponential speed-up over known classical algorithms for problems such as factoring integers~\cite{Sho94}, finding solutions to linear systems of equations~\cite{HHL} and simulating physical systems~\cite{Fey82,Llo96}. Quantum error-correction methods have been devised for preserving quantum information in the presence of noise~\cite{Sho95,CS,Ste96}, leading to the theoretical development of a fault-tolerant theory of quantum computing~\cite{AB-O,KLZ,Pre97}. Such a theory promises that quantum computation is possible in the presence of errors, provided the error rate is below a certain threshold value which depends on the particular coding scheme used as well as the error model. This potential has motivated much experimental research dedicated to building a functioning quantum information processor, with various proposals for possible implementations~\cite{CZ,CFH,LD,NPT}.

One of the main challenges in building a quantum information processor
is the non-scalability of \textit{completely} characterizing the noise affecting a quantum system via process tomography~\cite{CN,PCZ}.
A complete characterization of the noise is useful because it allows for the determination of good error-correction schemes, and thus the possibility of reliable transmission of quantum information. Since complete process tomography is infeasible for large systems, there is growing interest in scalable methods for \textit{partially} characterizing the noise affecting a quantum system~\cite{DCEL,EAZ,LLEC,ESMR,SMKE,BPP,SLCP,FL}. 

In Ref.~\cite{MGE} we provided a \emph{scalable} (in the number $n$ of qubits comprising the system) and robust method for benchmarking the full set of Clifford gates by a single parameter using randomization techniques. The concept of using randomization methods for benchmarking quantum gates, commonly called \textit{randomized benchmarking} (RB), was introduced previously in~\cite{EAZ,KLRB}. The simplicity of these protocols has motivated experimental implementations in atomic ions for different types of traps \cite{KLRB,Biercuk2009,BWC}, NMR \cite{RLL}, superconducting qubits \cite{CGT,Chow2010a}, and atoms in optical lattices \cite{Olmschenk2010}. Unfortunately there are several drawbacks to the methods of~\cite{EAZ,KLRB}. For instance~\cite{EAZ} assumes the highly idealized situation of the noise being independent of the chosen gate, in which case the fidelity decay curve averaged over randomly chosen unitaries takes the form of an exponential (in the sequence length). The protocol of~\cite{KLRB} is limited to the single-qubit case and fits the observed fidelity decay averaged over sequences of single-qubit gates (where each gate consists of a random generator of the Clifford group composed with a random Pauli operator) to an exponential. The decay rate is \emph{assumed} to provide an estimate of the average error probability per Clifford gate. However, conditions for when the assumption of an exponential decay is valid, specifically in the realistic case of gate-dependent and time-dependent noise, were not given. Such a set of conditions would be useful because it is easy to construct pathological examples where the estimated decay rate is not reliable. An unphysical but intuitively simple example is when the error is gate-dependent and equal to the exact inverse of the target gate. The error rate given by the protocol is always equal to zero however in actuality there is substantial error on each gate (see Sec. \ref{sec:Case}). Other important shortcomings of these previous RB protocols are that extensions to multi-qubit systems are either not scalable or not well understood, and it is unclear how to explicitly account for state preparation and measurement errors.

In this paper we give a full analysis of the scalable multi-qubit randomized benchmarking protocol for Clifford gates we proposed in~\cite{MGE} which overcomes the shortcomings described above. We note that since one ``gate" in the single-qubit protocol of~\cite{KLRB} consists of a random Clifford generator as well as a random Pauli operator, the cost of implementing a gate in this scheme is 2. In the single-qubit case, our RB scheme can be implemented by explicitly writing down the 24 elements of the Clifford group decomposed into a sequence of the same generators that are randomly applied in~\cite{KLRB}. The average number of generators in such a decomposition is 1.875 which implies that even for the single-qubit case our protocol takes no more time to implement than that of~\cite{KLRB}. Hence, since our protocol is scalable and produces an error-estimate which overcomes the various shortcomings listed above, it is reasonable to apply it over other existing schemes regardless of the number of qubits comprising the system.

We provide a detailed proof that our protocol requires at most $O\left(n^2\right)$ quantum gates, $O\left(n^4\right)$ cost in classical pre-processing (to select each gate-sequence), and a number of single-shot repetitions that is independent of $n$. As well, we give a thorough explanation of the perturbative expansion of the time and gate-dependent errors about the average error that leads to the fitting models for the observed fidelity decay. Our zeroth order model directly shows that for time-independent and gate-independent errors the fidelity decay is indeed modeled by an exponential decay, and the decay rate produces an estimate for the average error rate of the noise.

We derive the first order fitting model which takes into account the first-order correction terms in the perturbative expansion and provide a detailed explanation of the conditions for when this is a sufficient model of the fidelity decay curve. The fitting formula shows that gate-dependent errors can lead to a deviation from the exponential decay (defining a partial test for such effects in the noise), which was illustrated via numerical examples in~\cite{MGE}. State-preparation and measurement errors appear as independent fit parameters in the fitting models and we discuss when the protocol is robust against these errors. In the case of Pauli errors we give some novel preliminary results regarding the relationship between the benchmarking average error rate and the more common diamond norm error measure~\cite{AKN,KYV} used in fault-tolerant theory.

The paper is structured as follows: In section \ref{sec:Background} we discuss notation and background material. In section \ref{sec:Protocol} we discuss the proposed protocol and then in section \ref{sec:Fitting Models} we present the perturbative expansion and expressions for the zero'th and first order fitting models. Section \ref{sec:Neglectinghigherorders} provides a sufficient condition for neglecting higher order terms in the model as well as a simple case for when the benchmarking scheme fails. We also discuss when the protocol is robust against state preparation and measurement errors. Section \ref{sec:Normrelationships} discusses the relationship between the error rate given by the benchmarking scheme and other measures of error commonly used in quantum information. Section \ref{sec:Scalability} provides a detailed proof that our protocol is scalable in the number of qubits comprising the system, and a discussion with concluding remarks is contained in section \ref{sec:Discussion}.



\maketitle


\section{Background}\label{sec:Background}

Let us first set some notation. Suppose we have an $n$-qubit quantum system so that the Hilbert space $\Hilb$ representing the system has dimension $d=2^n$. Thus $\Hilb$ is isomorphic to $\mathbb{C}^d$ and both will generically refer to the Hilbert space of a $d$-dimensional quantum system throughout the presentation. The set of linear operators on $\Hilb$ will be denoted by $L\left(\Hilb\right)$. The set of pure states is represented by complex projective space $\pure$ and the set of all mixed states in $L\left(\Hilb\right)$, denoted by $\mathcal{D}(\Hilb)$, is given by the set of non-negative, trace-1 linear operators on $\Hilb$. Unless otherwise stated, we will only be concerned with quantum operations with the same input and output spaces. The set of linear superoperators mapping $L\left(\Hilb\right)$ into itself is denoted by $\mathcal{T}(\Hilb)$ with the set of quantum channels (completely positive, trace-preserving linear maps) contained in $\mathcal{T}(\Hilb)$ denoted by $\mathcal{S}(\Hilb)$.

There are various methods for quantifying the distance between quantum operations, we briefly describe those that will be of use to us. Good references for many of the topics in this section are~\cite{HJ,Wat05,Paul}.

 \subsection{Diamond Norm, Average Gate Fidelity and Minimum Gate Fidelity}\label{sec:norms}

One method of quantifying the distance between two linear superoperators $\Eop_1$, $\Eop_2 \in \mathcal{T}(\Hilb)$ is given by the diamond norm distance, $\|\Eop_1 - \Eop_2\|_{\diamond}$. The diamond norm of an arbitrary linear superoperator $\mathcal{R}: L\left(\mathbb{C}^m\right) \rightarrow L\left(\mathbb{C}^n\right)$ is defined as,

\begin{equation} 
\|\mathcal{R}\|_{\diamond} = \text{sup}_{k \in \mathbb{N}} \|\mathcal{R} \otimes \mathcal{I}_k\|_1 
\end{equation}

\noindent where $\| \: \|_1$ on superoperators is defined to be the $\infty$-norm induced by the trace norm $\| \: \|_1$ on $L\left(\mathbb{C}^m\right)$ and $L\left(\mathbb{C}^n\right)$. It is known that the supremum occurs for $k=m$ and so,

\begin{eqnarray}
\|\mathcal{R}\|_{\diamond} &=& \|\mathcal{R} \otimes \mathcal{I}_m\|_1 \nonumber \\
&=& \text{max}_{A: \|A\|_1 \leq 1}\| \mathcal{R} \otimes \mathcal{I}_m (A)\|_1 \label{eq:Diamond2}
\end{eqnarray}

\noindent where $A \in L\left(\mathbb{C}^m\otimes \mathbb{C}^m\right)$. Hence for $\Eop_1$, $\Eop_2 \in \mathcal{T}(\Hilb)$,

\begin{equation}
\|\Eop_1 - \Eop_2\|_{\diamond} = \|\left(\Eop_1 - \Eop_2\right) \otimes \mathcal{I}_d\|_{1}. 
\end{equation}

\noindent The diamond norm distance is commonly used in quantum information due to its operational meaning of being related to the optimal probability for distinguishing $\Eop_1$ and $\Eop_2$ using a binary outcome POVM and single input state (allowing for ancillas)~\cite{Sac05}.

Another method for quantifying the distance between linear superoperators is given by the $\| \: \|_{1\rightarrow 1}^H$ norm defined for linear superoperator $\mathcal{R}: L\left(\mathbb{C}^m\right) \rightarrow L\left(\mathbb{C}^n\right)$ as,

\begin{equation}\label{eq:onetooneH}
\|\Rop\|_{1\rightarrow 1}^H = \text{max}_{A : A=A^{\dagger}, \|A\|_1\leq 1} \|\Rop\left(A\right)\|_1 
\end{equation}

\noindent where $A \in L\left(\mathbb{C}^m\right)$. One can see that $\| \: \|_{1\rightarrow 1}^H$ is just $\| \: \|_{1}$ (which is also denoted $\| \: \|_{1\rightarrow 1}$) restricted to Hermitian inputs. This norm is less common in quantum information due to its lack of operational meaning, however it is a weaker measure of distance than the diamond norm since for any linear superoperator $\mathcal{R}: L\left(\mathbb{C}^m\right) \rightarrow L\left(\mathbb{C}^n\right)$, $\|\Rop\|_{1\rightarrow 1}^H \leq \|\Rop\|_{\diamond}$. This will be of much use to us later when we consider neglecting higher order effects in the benchmarking scheme.

A commonly used state-dependent measure for comparing quantum operations $\Eop_1$, $\Eop_2 \in \mathcal{S}(\Hilb)$ is given by the channel fidelity,

 \begin{eqnarray}
 \mathcal{F}_{\Eop_1,\Eop_2}(\rho) &=& F\left(\Eop_1(\rho),\Eop_2(\rho)\right) \nonumber \\
 &=& \left(\tr\sqrt{\sqrt{\Eop_1(\rho)} \Eop_2(\rho) \sqrt{\Eop_1(\rho)}}\right)^2 
 \end{eqnarray}

\noindent where ``$F$" refers to the usual fidelity between quantum states~\cite{NC}. In the case of a unitary operation $\mathcal{U}$, quantum operation $\Eop$, and restricting input states to $\pure$, the channel fidelity is called the gate fidelity. Explicitly, for $\phi \leftrightarrow |\phi\rangle \langle \phi | \in \pure$,

\begin{equation}
\mathcal{F}_{\Eop,\mathcal{U}}(\phi) = \text{tr}\left(\mathcal{U}(|\phi\rangle \langle \phi |)\Eop(|\phi\rangle \langle \phi |)\right),
\end{equation}

\noindent and defining $\Lambda = \mathcal{U}^{\dagger} \circ \Eop$ gives,

\begin{equation}
\mathcal{F}_{\Eop,\mathcal{U}}(\phi) = \mathcal{F}_{\Lambda,\mathcal{I}}(\phi) = \text{tr}\left(|\phi\rangle \langle \phi | \Lambda (|\phi\rangle \langle \phi |)\right). 
\end{equation}

\noindent The channel $\Lambda$ can be thought of as representing how much $\Eop$ deviates from $\mathcal{U}$ in that if $\Eop = \mathcal{U}$ then $\Lambda = \mathcal{I}$. The gate fidelity has many nice mathematical properties including a simple expression for the average over pure states, expressions for the variance in terms of various representations of $\Lambda$ and a concentration of measure phenomenon for large systems~\cite{Nie02,MBKE,MAG11}. The average gate fidelity is obtained by integrating $\mathcal{F}_{\Eop,\mathcal{U}}$ over $\pure$ using the Fubini-Study measure $\mu_{FS}$~\cite{BZ},

\begin{equation}
\overline{\mathcal{F}_{\Eop,\mathcal{U}}}=\overline{\mathcal{F}_{\Lambda,\mathcal{I}}}=\int_{\pure}\text{tr}\left(|\phi\rangle \langle \phi | \Lambda (|\phi\rangle \langle \phi |)\right)d\mu_{FS}(\phi).
\end{equation}

\noindent Taking the minimum of $\mathcal{F}_{\Eop_1,\Eop_2}$ over all mixed states $\rho$ produces a quantity $\mathcal{F}_{\Eop_1,\Eop_2}^{\text{min}}$ commonly called the minimum channel fidelity,

%

\begin{equation}
\mathcal{F}_{\Eop_1,\Eop_2}^{\text{min}} = \text{min}_{\rho}\mathcal{F}_{\Eop_1,\Eop_2} (\rho).\nonumber
\end{equation}

\noindent Note that by concavity of the fidelity, the minimum channel fidelity occurs at a pure state~\cite{NC}. In the case of the gate fidelity, the minimum is called the minimum gate fidelity.

In certain cases we will be concerned with how close $\Eop_1$ and $\Eop_2$ are in terms of the difference between the average fidelity of each channel. To this end we define,

\begin{gather}
\Delta F(\Eop_1,\Eop_2):=\left|    \overline{\mathcal{F}_{\Eop_1,\mathcal{I}}}   -   \overline{\mathcal{F}_{\Eop_2,\mathcal{I}}} \right|.
\end{gather}


Lastly, we note the following relationships between some of the distance measures defined above. First, for $\Eop_1$, $\Eop_2 \in \mathcal{S}(\Hilb)$ the following inequalities hold,

\begin{equation}\label{eq:ineqs}
\Delta F(\Eop_1,\Eop_2) \leq \|\Eop_1 - \Eop_2\|_{1 \rightarrow 1}^H \leq \|\Eop_1 - \Eop_2\|_{\diamond}.
 \end{equation}

\bigskip

\noindent where we recall the definition of $\| \: \|_{1\rightarrow 1}^H$ in Eq. (\ref{eq:onetooneH}. The second inequality is clear since,

\begin{equation}\label{eq:normineqs}
\|\Eop_1-\Eop_2\|_{1\rightarrow 1}^H \leq  \|\Eop_1-\Eop_2\|_{1}\leq \|\Eop_1-\Eop_2\|_{\diamond}.
\end{equation}

\noindent Now for the first inequality note that,

\begin{eqnarray}
\Delta F(\Eop_1,\Eop_2)&\leq& \text{max}_{|\phi\rangle} \left|\text{tr}\left(\left(\Eop_1-\Eop_2\right)\left(|\phi\rangle \langle \phi|\right)|\phi\rangle \langle \phi|\right)\right| \nonumber \\
&\leq&\text{max}_{|\phi\rangle} \left\|\left(\Eop_1-\Eop_2\right)\left(|\phi\rangle\langle \phi |\right)\right\|_{\infty}\nonumber \\
&=&\text{max}_{A:A=A^{\dagger}, \|A\|_1\leq 1}\left\|\left(\Eop_1-\Eop_2\right)\left(A\right)\right\|_{\infty} \nonumber \\
&=& \left\|\Eop_1-\Eop_2\right\|_{1\rightarrow \infty}^H 
\end{eqnarray}

\noindent where we note that since $\Eop_1$ and $\Eop_2$ are completely positive, $\Eop_1-\Eop_2$ is Hermiticity-preserving. Hence since $\left\|\Eop_1-\Eop_2\right\|_{1\rightarrow \infty}^H \leq \left\|\Eop_1-\Eop_2\right\|_{1\rightarrow 1}^H$ the inequalities in Eq. (\ref{eq:ineqs}) hold.

Next we show that for any quantum operations $\Eop_1$, $\Eop_2 \in \mathcal{S}(\Hilb)$,

\begin{equation}
\mathcal{F}_{\Eop_1,\Eop_2}^{\text{min}} \geq 1- \| \Eop_1 - \Eop_2\|_{\diamond}.
\end{equation}

\noindent We have that,

\begin{equation}
\| \Eop_1 - \Eop_2\|_{\diamond} = \text{max}_{|\psi\rangle \in \Hilb \otimes \Hilb} \| \Eop_1 \otimes \mathcal{I} (|\psi \rangle \langle \psi |) - \Eop_2 \otimes \mathcal{I} (|\psi \rangle \langle \psi |)\|_1.
\end{equation}

\noindent By the Fuchs-Van de Graaf inequalities~\cite{FvdG},

\begin{gather}
\| \Eop_1 \otimes \mathcal{I} (|\psi \rangle \langle \psi |) - \Eop_2 \otimes \mathcal{I} (|\psi \rangle \langle \psi |)\|_1 \geq \nonumber \\
1 - F(\Eop_1 \otimes \mathcal{I} (|\psi \rangle \langle \psi |),\Eop_2 \otimes \mathcal{I} (|\psi \rangle \langle \psi |)) 
\end{gather}

\noindent so,

\begin{gather}
\| \Eop_1 - \Eop_2\|_{\diamond} \nonumber \geq \\
\text{max}_{|\psi\rangle \in \Hilb \otimes \Hilb} \left[1 - F(\Eop_1 \otimes \mathcal{I} (|\psi \rangle \langle \psi |),\Eop_2 \otimes \mathcal{I} (|\psi \rangle \langle \psi |))\right] \nonumber \\
= 1-\text{min}_{|\psi\rangle \in \Hilb \otimes \Hilb}F(\Eop_1 \otimes \mathcal{I} (|\psi \rangle \langle \psi |),\Eop_2 \otimes \mathcal{I} (|\psi \rangle \langle \psi |)).
\end{gather}

Now we have,

\begin{gather}
\text{min}_{|\psi\rangle \in \Hilb \otimes \Hilb}F(\Eop_1 \otimes \mathcal{I} (|\psi \rangle \langle \psi |),\Eop_2 \otimes \mathcal{I} (|\psi \rangle \langle \psi |)) \leq \nonumber \\
\text{min}_{|\phi\rangle \in \Hilb}F(\Eop_1 (|\phi \rangle \langle \phi |),\Eop_2 (|\phi \rangle \langle \phi |))
\end{gather}

\noindent since

\begin{widetext}

\begin{gather}
\text{min}_{|\psi\rangle \in \Hilb \otimes \Hilb}F(\Eop_1 \otimes \mathcal{I} (|\psi \rangle \langle \psi |),\Eop_2 \otimes \mathcal{I} (|\psi \rangle \langle \psi |)) \leq \text{min}_{|\phi\rangle \in \Hilb} F(\Eop_1 \otimes \mathcal{I} (|\phi \rangle \langle \phi | \otimes |\phi \rangle \langle \phi |),\Eop_2 \otimes \mathcal{I} (|\phi \rangle \langle \phi | \otimes |\phi \rangle \langle \phi |)) \nonumber \\
= \text{min}_{|\phi\rangle \in \Hilb} \left(\text{tr} \sqrt{ \sqrt{ \Eop_1 (|\phi \rangle \langle \phi |) \otimes |\phi \rangle \langle \phi | } \left(\Eop_2 (|\phi \rangle \langle \phi |) \otimes |\phi \rangle \langle \phi | \right)\sqrt{\Eop_1 (|\phi \rangle \langle \phi |) \otimes |\phi \rangle \langle \phi |} }\right)^2 \nonumber \\
= \text{min}_{|\phi\rangle \in \Hilb} \left(\text{tr} \left( \sqrt{ \sqrt{\Eop_1 (|\phi \rangle \langle \phi |)} \left(\Eop_2 (|\phi \rangle \langle \phi |) \right)\sqrt{ \Eop_1 (|\phi \rangle \langle \phi |)}} \otimes |\phi \rangle \langle \phi | \right)\right)^2 \nonumber \\
= \text{min}_{|\phi\rangle \in \Hilb}F(\Eop_1 (|\phi \rangle \langle \phi |),\Eop_2 (|\phi \rangle \langle \phi |)).
\end{gather}

\end{widetext}

\noindent So,

\begin{gather}
\| \Eop_1 - \Eop_2\|_{\diamond} \geq 1 - \text{min}_{|\phi\rangle \in \Hilb}F(\Eop_1 (|\phi \rangle \langle \phi |),\Eop_2 (|\phi \rangle \langle \phi |)).
\end{gather}

Now by concavity,

%

\begin{gather}
\mathcal{F}_{\Eop_1,\Eop_2}^{\text{min}} = \text{min}_{|\phi\rangle \in \Hilb}F(\Eop_1 (|\phi \rangle \langle \phi |),\Eop_2 (|\phi \rangle \langle \phi |))
\end{gather}

\noindent and so,

\begin{gather}
\mathcal{F}_{\Eop_1,\Eop_2}^{\text{min}} \geq 1 - \| \Eop_1 - \Eop_2\|_{\diamond}. \label{eq:minfidtodiamond}
\end{gather}

%

\subsection{The Clifford Group and t-Designs}\label{sec:Cliffordgroup}

The Clifford group on $n$ qubits, denoted $\Clif$, is defined as the normalizer of the Pauli group $\mathcal{P}_n$ and is generated by the phase (S), Hadamard (H) and controlled-NOT (CNOT) gates. $\Clif$  plays an important role in many areas of quantum information such as universality~\cite{BMPRV}, stabilizer code theory/fault-tolerance~\cite{Got97} and noise estimation~\cite{DCEL}.

One extremely useful property of $\Clif$, especially for noise estimation, is that the uniform probability distribution over $\Clif$ comprises a unitary 2-design~\cite{DCEL}. A unitary t-design is defined as follows,

\begin{definition} Unitary t-Design

A unitary t-design is a discrete random variable $\{(q_1,U_1),...,(q_K,U_K)\}$, with each $U_i \in U(d)$, such that for every homogeneous complex-valued polynomial p in $2d^2$ indeterminates of degree (s,s) less than or equal to (t,t),

\begin{equation}
\frac{1}{K} \sum _{j=1}^K p(U_j) = \int _{U(d)}p(U)dU.
\end{equation}

\end{definition}

\noindent The integral is taken with respect to the Haar measure on $ U(d)$. Here $p(U)$ is defined to be the evaluation of $p$ at the $2d^2$ values consisting of the $d^2$ matrix entries of $U$ as well as the $d^2$ complex conjugates of these matrix entries. In the case $t=2$ the above reduces to a ``twirling"~\cite{BDSW} condition,

\begin{equation}
\sum _{j=1}^K q_j\left(U_j\Lambda \left(U_j^{\dagger} \rho U_j\right) U_j^{\dagger}\right) = \int _{U(d)}\left(U\Lambda \left(U^{\dagger} \rho U\right) U^{\dagger}\right)dU 
\end{equation}

\noindent being satisfied for any quantum channel $\Lambda$ and any state $\rho$ ~\cite{DCEL}. Since a uniform probability distribution on $\Clif$ forms a 2-design, if $\Clif = \{\mathcal{C}_j : j \in \mathcal{K} = \{1,...,\left|\Clif\right|\}\}$ then,

\begin{eqnarray}
\mathcal{W}(\Lambda)(\rho) &:=& \frac{1}{\left|\Clif\right|}\sum _{j=1}^{\left|\Clif\right|} \left(C_j\Lambda \left(C_j^{\dagger} \rho C_j\right) C_j^{\dagger}\right) \nonumber \\
&=& \int _{U(d)}\left(U\Lambda \left(U^{\dagger} \rho U\right) U^{\dagger}\right)dU.
\end{eqnarray}

As shown in~\cite{Nie02,EAZ}, $\int _{U(d)}\left(U\Lambda \left(U^{\dagger} \rho U\right) U^{\dagger}\right)dU$ produces the unique depolarizing channel $\Lambda_{d}$ with the same average fidelity as $\Lambda$. Hence if $\overline{\mathcal{F}_{\Lambda,\mathcal{I}}}$ is the average fidelity of $\Lambda$, and $\Lambda_{d}$ is given by

\begin{equation}
\Lambda_{d}(\rho)=p\rho + (1-p)\frac{\Id}{d}
\end{equation}

\noindent then,

\begin{equation}
\overline{\mathcal{F}_{\Lambda,\mathcal{I}}}=p+\frac{(1-p)}{d}.
\end{equation}

\noindent Thus twirling a quantum operation over the Clifford group produces a depolarizing channel and the average fidelity is invariant under the twirling operation.

In Sec. \ref{sec:RandomizedBenchmarking} we will be concerned with compositions of both gate-independent and gate-dependent twirls. In the gate-independent case, the sequence of twirls of $\Lambda$ of length k, $\mathcal{W}(\Lambda)^{k}$,
can be re-written as the k-fold composition of $\Lambda_{\text{d}}$ with itself. Using the above representation of $\Lambda_{\text{d}}$ we get,

\begin{equation}
\mathcal{W}(\Lambda)^{k}(\rho)=p^k\rho + (1-p^k)\frac{\Id}{d}.
\end{equation}

\noindent Therefore the average fidelity decreases exponentially to $\frac{1}{d}$ since,

\begin{equation}
\overline{\mathcal{F}_{\Lambda_{d}^k,\mathcal{I}}}=p^k+\frac{(1-p^k)}{d}.
\end{equation}

We can also write the average fidelity of $\Lambda$ in terms of its $\chi$-matrix~\cite{CN}. The $\chi$-matrix is an important (basis-dependent) object in experimental quantum information as it is directly related to practical methods in process tomography. The $\chi$-matrix is obtained by expanding the Kraus operators $\{A_k\}$ of $\Lambda$ with respect to a particular basis of $L\left(\mathbbm{C}^d\right)$, which is most often chosen to be the Pauli basis $\{P_j\}_{j=0}^{d^2-1}$ ($P_0=\openone$). This gives,

\begin{equation}
\Lambda(\rho)=\sum_k A_k\rho A_k^{\dagger} = \sum_{i,j}\chi_{i,j}P_i\rho P_j
\end{equation}

\noindent and so a complete description for $\Lambda$ can be given by estimating the entries of $\chi$. As shown in~\cite{CN},

\begin{equation}
\overline{\mathcal{F}_{\Lambda,\mathcal{I}}}=\frac{\chi_{0,0}d+1}{d+1}
\end{equation}

\noindent which gives,

\begin{equation}
\chi_{0,0}=p\left(1-\frac{1}{d^2}\right)+\frac{1}{d^2}=\frac{\overline{\mathcal{F}_{\Lambda,\mathcal{I}}}(d+1)-1}{d}.
\end{equation}

\noindent Therefore the $(0,0)$ entry of the $\chi$-matrix for a quantum operation with respect to the Pauli basis is invariant under twirling over a 2-design. Moreover $\chi_{0,0}$ for $\Lambda_{dep}^k$ decreases to $\frac{1}{d^2}$ exponentially in k.

\section{Randomized Benchmarking}\label{sec:RandomizedBenchmarking}

In this section we present both the protocol and a full derivation of the fitting models for randomized benchmarking that were given in~\cite{MGE}. First, we set some notation and make various definitions that will be used throughout the presentation. 

Denote the elements of $\Clif$ by $\mathcal{C}_i$ and the maximum sequence length of applying Clifford gates by $M$.  Suppose that the actual implementation of $\mathcal{C}_{i}$ at time j ($1\leq j \leq M$) results in the map $\Eop _{i,j}$ with $\Eop_{i,j} = \Lambda_{i,j} \circ \mathcal{C}_{i}$ for some error map $\Lambda_{i,j}$. Hence to each Clifford $\mathcal{C}_i$ we associate a sequence $\Lambda_{i,1},...,\Lambda_{i,M}$ which represents the time-dependent noise operators affecting $\mathcal{C}_i$.  We define the average error operator as follows,

\begin{definition}

Average Error Operator

The average error operator affecting the gates in $\Clif$ is given by,

\begin{equation}
\Lambda= \frac{1}{M\left|\Clif\right|}\sum_j\sum_i\Lambda_{i,j}.
\end{equation}

\end{definition}



Consider the twirl of the average error operator over $\Clif$. As discussed in Sec.(\ref{sec:Cliffordgroup}) this produces a depolarized channel $\Lambda_{\text{d}}$,
\begin{equation}
\Lambda_{\text{d}}(\rho) = \frac{1}{ \left| \Clif \right| } \sum_i C_i^\dagger \circ \Lambda_{\mathrm{ave}} \circ  C_i \; (\rho )   = p \rho + (1-p) \frac{\Id}{d}.
\end{equation}
Recall from Sec. (\ref{sec:Cliffordgroup}) that the average fidelity of $\Lambda$, denoted $F_{\mathrm{ave}}$, is invariant under Clifford twirling and so,

\begin{equation}
F_{\mathrm{ave}} = p+\frac{1-p}{d}.
\end{equation}

\noindent We now define the average error rate of the set of Clifford gates as follows:

\begin{definition}

Average Error Rate


The average error rate, $r$, of the Clifford gates used in a quantum computation is defined to be,

\begin{equation}\label{eq:rate}
r=1-F_{\mathrm{ave}}=1-\left(p+\frac{1-p}{d}\right) = \frac{(d-1)(1-p)}{d}.
\end{equation}

\end{definition}

It is important to note that $r$ defined above should not be confused with the ``error rate", $r_{\mathcal{P}}$, of a Pauli channel $\mathcal{P}$. For Pauli channel $\mathcal{P}$, $r_{\mathcal{P}}$ is defined to be the probability that a non-identity Pauli operator is applied to the input state. Conditioning on a non-identity Pauli being applied, there is still a non-zero probability of the input state being unchanged. Subtracting this probability out gives our defined parameter $r$ for $\mathcal{P}$ which is commonly called the ``infidelity" of $\mathcal{P}$. One can show that $r$ and $r_{\mathcal{P}}$ are related via $r_{\mathcal{P}}=\frac{(d+1)r}{d}$. Following the terminology set in~\cite{KLRB} we will call $r$ the (average) error-rate of $\Lambda$ and note that in the case where $\Lambda$ is a Pauli channel, $r$ is equal to the infidelity of $\Lambda$.

The parameter $r$ is the figure of merit we want to be able to estimate experimentally. One can estimate $p$ directly using any of standard process tomography~\cite{CN}, ancilla-assisted/entanglement-assisted process tomography~\cite{ABJ} or Monte-Carlo methods~\cite{SLCP,FL}. The tomography based schemes suffer from the unrealistic assumptions of negligible state-preparation and measurement errors, and clean ancillary states/operations. These schemes also require exponential time resources in $n$ making them infeasible for even relatively small numbers of qubits. The Monte-Carlo methods also have the drawback of assuming negligible state-preparation and measurement errors. The advantages of these methods are that the average fidelity of each gate can be estimated and the scheme is efficient in $n$.


The experimentally relevant challenge therefore is to estimate $p$ while relaxing the assumptions on state preparation, measurement and ancillary states/processes. Ideally, such a method should also scale efficiently with the number of qubits. As we show below, such an estimate can be obtained through benchmarking the performance of random circuits.


\subsection{Protocol}\label{sec:Protocol}

For a fixed sequence length $m \leq M-1$, the benchmarking protocol consists of choosing $K_m$ sequences of independent and identically distributed uniformly random Clifford elements and calculating the fidelity of the average of the $K_m$ sequences. One repeats this procedure for different values of $m$ and fits the fidelity decay curve to the models we derive below. More precisely, the protocol is as follows,

\bigskip

\noindent Fix an initial state $|\psi\rangle$ and perform the following steps:

\bigskip

\emph{Step 1}. Fix $m \leq M-1$ and generate $K_m$ sequences consisting of $m+1$ quantum operations. The first $m$ operations are chosen uniformly at random from $\Clif$ and the $m+1$'th operation is uniquely determined as the inverse gate of the composition of the first $m$. By assumption each operation $\mathcal{C}_{i_j}$ is allowed to have some error, represented by $\Lambda_{i_{j},j}$, and each sequence can be modelled by the operation,

\begin{equation}\label{eq:seq}
\mathcal{S}_\mathbf{i_m} = \bigcirc_{j=1}^{m+1}\left(\Lambda_{i_{j},j}\circ\mathcal{C}_{i_j}\right),
\end{equation}

\noindent where $\mathbf{i_m}$ is the $m$-tuple $(i_1,...,i_m)$ (which we sometimes also denote by $\vec{i_m}$) and $i_{m+1}$ is uniquely determined by $\mathbf{i_m}$.


\bigskip

\emph{Step 2}. For each of the $K_m$ sequences, measure the survival probability $\mathrm{Tr} [ E_\psi \mathcal{S}_\mathbf{i_m} (\rho_\psi)]$. Here $\rho_\psi$ is a quantuml state that takes into account errors in preparing $|\psi\rangle \langle \psi |$ and $E_\psi$ is the POVM element that takes into account measurement errors. In the ideal (noise-free) case $\rho_{\psi}=E_{\psi}=\kett\braa$.

\bigskip

\emph{Step 3}. Average over the $K_m$ random realizations to find the averaged sequence fidelity,
\begin{equation}\label{eq:seqfid}
F_\mathrm{seq}(m,\psi) =  \mathrm{Tr} [ E_\psi \mathcal{S}_{K_m}  ( \rho_\psi)],
\end{equation} where
\begin{equation}\label{eq:aveS}
\mathcal{S}_{K_m} = \frac{1}{ K_m } \sum_{\mathbf{i_m}} \mathcal{S}_\mathbf{i_m}
\end{equation} is the average sequence operation.

\emph{Step 4}. Repeat Steps 1 through 3 for different values of $m$ and fit the results for the averaged sequence fidelity (defined in Eq.~\eqref{eq:seqfid}) to the model
\begin{equation}\label{eq:model}
\Fgone =A_1 p^{m}  +B_1+C_1  (m-1)(q-p^2) p^{m-2}
\end{equation} derived below. The coefficients $A_1$, $B_1$, and $C_1$ absorb the state preparation and measurement errors as well as the error on the final gate. The difference $q-p^2$ is a measure of the degree of gate-dependence in the errors, and $p$ determines the average error-rate $r$ according to the relation given by Eq. (\ref{eq:rate}). In the case of gate-independent and time-independent errors the results will fit the simpler model

\begin{equation} \label{eq:zerothorder}
\Fgzero = A_0 p^{m} + B_0
\end{equation}

also derived below, where $A_0$ and $B_0$ absorb state preparation and measurement errors as well as the error on the final gate.

We note that for each $m$, in the limit of $K_m \rightarrow \infty$, $F_\mathrm{seq}(m,\psi)$ converges to the exact (uniform) average, $\Fg$, over all sequences,

\begin{eqnarray}
\Fg&=& \mathrm{Tr} [ E_\psi \mathcal{S}_m  ( \rho_\psi) ] \nonumber \\ 
\end{eqnarray}

\noindent where we define the exact average of the sequences to be,

\begin{equation}
\mathcal{S}_m=  \frac{1}{\left|\Clif\right|^m}\sum_{\left(i_1,...,i_m\right)}\Lambda_{i_{m+1},m+1} \circ \mathcal{C}_{i_{m+1}} \circ ...  \circ \Lambda_{i_1,1}  \circ \mathcal{C}_{i_1}. \label{eq:Smdef}
\end{equation}

\noindent Hence the fitting functions by which we model the behavior of $F_\mathrm{seq}(m,\psi)$ are derived in terms of $\Fg$ (see Sec.~\ref{sec:Fitting Models}). Note that since $\Fg$ is the uniform average over all sequences we can sum over each index independently,

\begin{widetext}

\begin{equation}
\Fg= \frac{1}{\left|\Clif\right|^m}\sum_{i_1,...,i_m}\tr\left(\Lambda_{i_{m+1},m+1} \circ \mathcal{C}_{i_{m+1}} \circ \Lambda_{i_m,m} \circ \mathcal{C}_{i_m}\circ  ...  \circ \Lambda_{i_1,1}  \circ \mathcal{C}_{i_1} ( \rho_{\psi} ) E_{\psi}\right).\label{eq:avgdepseq}
\end{equation}

\end{widetext}

In order to prepare for the next section where we derive the above fitting models, we write $\Fg$ in a more intuitive form. We first re-write $\Lambda_{i_{m+1},m+1} \circ \mathcal{C}_{i_{m+1}} \circ \Lambda_{i_{m},m} \circ \mathcal{C}_{i_m} \circ ...\circ  \Lambda_{i_{1},1} \circ \mathcal{C}_{i_1}$ by inductively defining new uniformly random gates from the Clifford group in the following manner:

\bigskip

1. Define $\mathcal{D}_{i_1}= \mathcal{C}_{i_1}$.

\bigskip

2. Define $\mathcal{D}_{i_2}$ uniquely by the equation $\mathcal{C}_{i_2} = \mathcal{D}_{i_2} \circ \mathcal{D}_{i_1}^{\dagger}$, ie. $\mathcal{D}_{i_2}=\mathcal{C}_{i_2} \circ \mathcal{C}_{i_1}=\bigcirc_{s=1}^2\mathcal{C}_{i_s}$.

\bigskip

3. In general, for j $\in \{2,...,m\}$, if $\mathcal{C}_{i_1}$,...,$\mathcal{C}_{i_j}$ and ${\mathcal{D}_{i_1}}$,...,${\mathcal{D}_{i_j}}$ have been chosen, define ${\mathcal{D}_{i_{j+1}}}$ uniquely by the equation $\mathcal{C}_{i_{j+1}} = {\mathcal{D}_{i_{j+1}}} \circ {\mathcal{D}_{i_j}}^{\dagger}$, ie. 

\begin{equation}
{\mathcal{D}_{i_{j+1}}} = \mathcal{C}_{i_{j+1}} \circ ... \circ \mathcal{C}_{i_1}=\bigcirc_{s=1}^{j+1}\mathcal{C}_{i_s}.
\end{equation}

\bigskip

\noindent Note that if $j \neq k$, $\mathcal{C}_{i_j}$ and $\mathcal{C}_{i_k}$ are independent and so since the Clifford elements form a group, for each $j=2,...,m+1$, ${\mathcal{D}_{i_j}}$ is independent of ${\mathcal{D}_{i_{j-1}}}$. As well, summing over each $i_j$ index runs over every Clifford element once and only once in ${\mathcal{D}_{i_j}}$. 

We have created a new sequence $\left({\mathcal{D}_{i_1}},...,{\mathcal{D}_{i_m}}\right)$ from $\left(\mathcal{C}_{i_1},...,\mathcal{C}_{i_m}\right)$ uniquely so that

\begin{eqnarray}
\mathcal{S}_{\vec{i_m}} &=& \Lambda_{i_{m+1},m+1} \circ \mathcal{C}_{i_{m+1}} \circ \Lambda_{i_{m},m} \circ \mathcal{C}_{i_m} \circ ...\circ  \Lambda_{i_{1},1} \circ \mathcal{C}_{i_1} \nonumber \\
&=& \Lambda_{i_{m+1},m+1} \circ {\mathcal{D}_{i_{m+1}}}\circ {\mathcal{D}_{i_{m}}}^{\dagger} \circ  \Lambda_{i_{m},m}\circ {\mathcal{D}_{i_{m}}} \circ ... \nonumber \\
& \: & \circ {\mathcal{D}_{i_1}}^{\dagger} \circ  \Lambda_{i_{1},1} \circ {\mathcal{D}_{i_1}}.
\end{eqnarray}

\noindent Since $\mathcal{C}_{i_{m+1}}=\mathcal{C}_{i_1}^{\dagger}\circ...\circ \mathcal{C}_{i_m}^{\dagger}$ and ${\mathcal{D}_{i_{m+1}}}=\mathcal{C}_{i_{m+1}} \circ...\circ \mathcal{C}_{i_1}$,

\begin{equation}
{\mathcal{D}_{i_{m+1}}} = \Id.
\end{equation}

\noindent Hence the m+1'th gate is decoupled from the rest of the sequence and we have

\begin{eqnarray}
\mathcal{S}_{\vec{i_m}} &=& \Lambda_{i_{m+1},m+1} \circ \mathcal{C}_{i_{m+1}} \circ \Lambda_{i_{m},m} \circ \mathcal{C}_{i_m} \circ ...\circ  \Lambda_{i_{1},1} \circ \mathcal{C}_{i_1} \nonumber \\
&=& \Lambda_{i_{m+1},m+1} \circ {\mathcal{D}_{i_{m}}}^{\dagger} \circ  \Lambda_{i_{m},m}\circ {\mathcal{D}_{i_{m}}} \circ ...  \nonumber \\
& \: & \circ {\mathcal{D}_{i_1}}^{\dagger} \circ  \Lambda_{i_{1},1} \circ {\mathcal{D}_{i_1}}. 
\end{eqnarray}


\subsection{Perturbative Expansion and the Fitting Models}\label{sec:Fitting Models}

We would like to develop fitting models for $\Fg$ where the most general noise model allows for the noise to depend upon both the set of gates in $\Clif$ and time.  We can estimate the behavior of $\Fg$ by considering a perturbative expansion of each $\Lambda_{i,j}$ about the average $\Lambda$. We quantify the difference between $\Lambda_{i,j}$ and $\Lambda$ by defining for all i, j,

\begin{equation}
\delta \Lambda_{i,j} = \Lambda_{i,j} - \Lambda.
\end{equation}

\noindent Our approach will be valid provided $\delta \Lambda_{i,j}$ is a small perturbation from $\Lambda$ in a sense to be made precise later. Note that each $\delta \Lambda_{i,j}$ is a Hermiticity-preserving, trace-annihilating linear superoperator. Under the above conditions this approach will allow for fitting the experimental fidelity decay sequence to a model with fit parameters that determine not only the average error per gate but also the separate contribution from the combined effects of state preparation and measurement errors. In the limit of multiple qubits and very precise control weaker forms of twirling may permit even more detailed modeling of the noise.

Using the change of variables ${\mathcal{D}_{i_j}}=\bigcirc_{s=1}^j \mathcal{C}_{i_s} $ described above and expanding to first order we get,

\begin{widetext}

\begin{eqnarray}
\mathcal{S}_{\vec{i_m}} &\equiv& \Lambda_{i_{m+1},m+1} \circ \mathcal{C}_{i_{m+1}} \circ ...\circ \Lambda_{i_j,j}  \circ \mathcal{C}_{i_j} \circ ...  \circ \Lambda_{i_1,1}  \circ \mathcal{C}_{i_1}   \nonumber \\
 &=& \Lambda_{i_{m+1},m+1} \circ {\mathcal{D}_{i_m}}^{\dagger} \circ \Lambda_{i_m,m} \circ {\mathcal{D}_{i_{m}}} \circ ... \circ {\mathcal{D}_{i_1}}^{\dagger} \circ \Lambda_{i_1,1}  \circ {\mathcal{D}_{i_1}}
\nonumber \\
 &=& \: \: \: \: \: \Lambda \circ {\mathcal{D}_{i_m}}^{\dagger} \circ \Lambda \circ {\mathcal{D}_{i_{m}}} \circ ... \circ {\mathcal{D}_{i_1}}^{\dagger} \circ \Lambda  \circ {\mathcal{D}_{i_1}}  + \: \delta \Lambda_{i_{m+1},m+1} \circ \left({\mathcal{D}_{i_m}}^{\dagger} \circ \Lambda \circ {\mathcal{D}_{i_{m}}}\right) \circ ... \circ \left({\mathcal{D}_{i_1}}^{\dagger} \circ \Lambda  \circ {\mathcal{D}_{i_1}}\right) \nonumber \\
 &\: \: & +...+  \Lambda  \circ \left({\mathcal{D}_{i_m}}^{\dagger} \circ \Lambda \circ {\mathcal{D}_{i_{m}}} \right)\circ ...   \circ \left({\mathcal{D}_{i_j}}^{\dagger} \circ \delta \Lambda_{i_j,j}   \circ {\mathcal{D}_{i_j}}\right) \circ ... \circ \left({\mathcal{D}_{i_1}}^{\dagger} \circ \Lambda \circ {\mathcal{D}_{i_1}}\right)\nonumber \\
 &\: \: & +...+ \Lambda  \circ \left({\mathcal{D}_{i_m}}^{\dagger} \circ \Lambda \circ {\mathcal{D}_{i_{m}}} \right)\circ ... \circ \left({\mathcal{D}_{i_1}}^{\dagger} \circ \delta\Lambda_{i_1,1}   \circ {\mathcal{D}_{i_1}}\right) + O(\delta \Lambda_{i_j,j}^2). \label{eq:expansion}
\end{eqnarray}


\noindent We define

\begin{gather}
\mathcal{S}_{\vec{i_m}}^{(0)} := \Lambda \circ {\mathcal{D}_{i_m}}^{\dagger} \circ \Lambda \circ {\mathcal{D}_{i_{m}}} \circ ... \circ {\mathcal{D}_{i_1}}^{\dagger} \circ \Lambda  \circ {\mathcal{D}_{i_1}}, 
\end{gather}


\begin{eqnarray}
\left(\mathcal{S}_{\vec{i_m}}^{(1)}\right) &:=& \: \delta \Lambda_{i_{m+1},m+1} \circ \left({\mathcal{D}_{i_m}}^{\dagger} \circ \Lambda \circ {\mathcal{D}_{i_{m}}}\right) \circ ... \circ \left({\mathcal{D}_{i_1}}^{\dagger} \circ \Lambda  \circ {\mathcal{D}_{i_1}}\right) \nonumber \\
 &\: \: & +...+  \Lambda  \circ \left({\mathcal{D}_{i_m}}^{\dagger} \circ \Lambda \circ {\mathcal{D}_{i_{m}}} \right)\circ ...   \circ \left({\mathcal{D}_{i_j}}^{\dagger} \circ \delta \Lambda_{i_j,j}   \circ {\mathcal{D}_{i_j}}\right) \circ ... \circ \left({\mathcal{D}_{i_1}}^{\dagger} \circ \Lambda \circ {\mathcal{D}_{i_1}}\right)\nonumber \\
 &\: \: & +...+ \Lambda  \circ \left({\mathcal{D}_{i_m}}^{\dagger} \circ \Lambda \circ {\mathcal{D}_{i_{m}}} \right)\circ ... \circ \left({\mathcal{D}_{i_1}}^{\dagger} \circ \delta\Lambda_{i_1,1}   \circ {\mathcal{D}_{i_1}}\right) \label{eq:firstorderexpansion}
\end{eqnarray}

\end{widetext}

\noindent and so on for higher order perturbation terms. As well, recalling the definition of $\mathcal{S}_{m}$ in Eq. (\ref{eq:Smdef}),  we define for each order $k$,

\begin{gather}
\mathcal{S}_m^{(k)} := \frac{1}{\left|\Clif\right|^m} \sum_{i_1,...,i_m} \mathcal{S}_{\vec{i_m}}^{(k)}
\end{gather}

\noindent and

\begin{gather}
\Fgk := \tr\left[\left(\sum_{j=0}^k\mathcal{S}_m^{(j)}\right)( \rho_{\psi} ) E_{\psi}\right]
\end{gather}

\noindent so that,

\begin{equation}
\mathcal{S}_m=\sum_{k=0}^{m+1}\mathcal{S}_m^{(k)},
\end{equation}

\noindent and

\begin{equation}
\Fg=\mathcal{F}_g^{(m+1)}(m,|\psi\rangle) = \tr\left[\left(\sum_{j=0}^{m+1}\mathcal{S}_m^{(j)}\right)(\rho_{\psi})E_{\psi}\right].
\end{equation}

\subsubsection{Zeroth Order Model}

First, we look at the zeroth order fitting model $\Fgzero$ and note that $\Fgzero$ is exact in the case that the noise is independent of both the gate chosen and time, ie. $\Lambda_{i_j,j} = \Lambda$. By independence of the $\mathcal{D}_{i_j}$ and the fact that averaging over the ensemble of realizations produces independent twirls which depolarize m factors of $\Lambda$ (see Sec. (\ref{sec:Cliffordgroup})) we get,

\begin{gather} \label{eq:Smzero}
S_m^{(0)} = \Lambda \circ \Lambda_d \circ ... \circ \Lambda_d = \Lambda \circ \left(\bigcirc_{j=1}^m \Lambda_d\right).
\end{gather}

\noindent Thus,

\begin{eqnarray}
\Fgzero&=&\text{tr}\left(S_m^{(0)}( \rho_{\psi} ) E_{\psi}\right)\nonumber \\
 &=& \tr\left(\Lambda( \rho_{\psi} ) E_{\psi}\right)p^m \nonumber \\
 & \: \: & + \tr\left(\Lambda\left(\frac{\Id}{d}\right)E_{\psi}\right)\left(1-p^m\right)\nonumber \\
&=& A_0p^m + B_0\label{eq:zerothorder}
\end{eqnarray}

\noindent where

\begin{equation}\label{eq:A0def}
A_0:=\mathrm{Tr}\left[E_\psi \Lambda\left(\rho_\psi-\frac{\Id}{d}\right)\right]
\end{equation}

\noindent and

\begin{equation}\label{eq:B0def}
B_0:=\mathrm{Tr}\left[E_\psi\Lambda\left(\frac{\Id}{d}\right)\right].
\end{equation}

\noindent Hence, assuming the simplest (ideal) scenario where the noise operator at each step is independent of the applied gate (and is also time-invariant), $\Fg = \Fgzero$ decays exponentially in $p$.

\subsubsection{First Order Model}

\noindent To find $\Fgone$ we note that in the definition of $S_{\vec{i_m}}^{(1)}$ given by Eq. (\ref{eq:firstorderexpansion}) there are $m+1\choose{1}$$=m+1$ first-order perturbation terms which contain the gate dependence. First, we consider the $m-1$ terms with $ j \in \{2,...,m\}$. For each such $j$, averaging over the $ \{i_1 ... i_m \}$ gives a term of the form,

\begin{gather}
\frac{1}{\left|\Clif\right|^m}\sum_{i_1 ... i_m}  \Lambda  \circ \left({\mathcal{D}_{i_m}}^{\dagger} \circ \Lambda \circ {\mathcal{D}_{i_{m}}} \right)\circ ... \nonumber \\
  \circ \left({\mathcal{D}_{i_j}}^{\dagger} \circ \delta \Lambda_{i_j,j}   \circ {\mathcal{D}_{i_j}}\right) \circ \left({\mathcal{D}_{i_{j-1}}}^{\dagger} \circ \Lambda \circ {\mathcal{D}_{i_{j-1}}}\right) \circ ... \nonumber \\
   \circ \left({\mathcal{D}_{i_1}}^{\dagger} \circ \Lambda \circ {\mathcal{D}_{i_1}}\right). 
\end{gather}

\noindent For these $m-1$ terms the main trick is to realize that we can re-expand ${\mathcal{D}_{i_j}} = \mathcal{C}_{i_j} \circ \mathcal{D}_{i_{j-1}} $ in order to depolarize the unitarily rotated perturbation  $ \mathcal{C}_{i_j}^{\dagger} \Lambda_{i_j,j} \mathcal{C}_{i_j} $ with the \emph{twirling} operation $  \frac{1}{\left|\Clif\right|}\sum_{i_{j-1}}  \mathcal{D}_{i_{j-1}}^{\dagger} \cdot \mathcal{D}_{i_{j-1}}$ because the sums are independent. More precisely, the above can be written as,

\begin{widetext}

\begin{gather}
\Lambda  \circ \Lambda_d^{m-j} \circ  \left[  \frac{1}{\left|\Clif\right|^2}\sum_{i_{j-1}, i_j}  {\mathcal{D}_{i_{j-1}}}^{\dagger} \circ \mathcal{C}_{i_j}^{\dagger} \circ \delta\Lambda_{i_j,j} \circ \mathcal{C}_{i_j}  \circ \Lambda \circ {\mathcal{D}_{i_{j-1}}} \right] \circ \left[\sum_{i_{j-2},...,i_1} \left({\mathcal{D}_{i_{j-2}}}^{\dagger} \circ \Lambda \circ {\mathcal{D}_{i_{j-2}}}\right) \circ ... \circ \left({\mathcal{D}_{i_1}}^{\dagger} \circ \Lambda \circ {\mathcal{D}_{i_1}}\right) \right]\nonumber \\
 = \Lambda  \circ \Lambda_d^{m-j} \circ
\left( \left(\mathcal{Q}_{j}\circ \Lambda\right)_d - \Lambda_d^2  \right) \circ  \Lambda_d^{j-2},
\end{gather}

\end{widetext}

\noindent where $\mathcal{Q}_{j}:=\frac{1}{\left|\Clif\right|} \sum_{i}   \mathcal{C}_i^{\dagger} \circ \Lambda_{i,j} \circ \mathcal{C}_i$ and the subscript d represents the depolarization of the operator within brackets. Using the fact that depolarizing channels commute we get,

\begin{gather} 
\Lambda  \circ \Lambda_d^{m-j} \circ \left( \left(\mathcal{Q}_{j} \circ \Lambda\right)_d - \Lambda_d^2  \right) \circ  \Lambda_d^{j-2} \nonumber \\
=  \Lambda  \circ \left( \left(\mathcal{Q}_{j} \circ \Lambda\right)_d - \Lambda_d^2  \right) \circ  \Lambda_d^{m-2}. \label{eq:exp1}
\end{gather}

For the term with $j =1$, averaging over $i_1,...,i_m$ gives a term of the form,

\begin{equation}\label{eq:exp2}
\Lambda  \circ \Lambda_d^{m-1} \circ  \frac{1}{\left|\Clif\right|} \sum_{i_1} {\mathcal{D}_{i_1}}^{\dagger} \circ \delta \Lambda_{i_1,1}   \circ {\mathcal{D}_{i_1}}
= \Lambda \circ \Lambda_d^{m-1} \circ (\mathcal{Q}_{1}  - \Lambda_d),
\end{equation}

where

\begin{eqnarray}
 \mathcal{Q}_{1} &:=& \frac{1}{\left|\Clif\right|} \sum_{i_1} \left(\mathcal{D}_{i_1}^{\dagger} \circ \Lambda_{i_1,1} \circ \mathcal{D}_{i_1} \right) \nonumber \\
 &=& \frac{1}{\left|\Clif\right|} \sum_{i} \left(\mathcal{C}_i^{\dagger} \circ \Lambda_{i,1} \circ \mathcal{C}_i \right).
 \end{eqnarray}

Lastly for the term with $j=m+1$, averaging gives,

\begin{widetext}

\begin{gather}
\frac{1}{\left|\Clif\right|^m}\sum_{i_1 ... i_m}  \delta\Lambda_{i_{m+1},m+1} \circ \left({\mathcal{D}_{i_m}}^{\dagger} \circ \Lambda \circ {\mathcal{D}_{i_{m}}} \right)\circ ... \circ \left({\mathcal{D}_{i_1}}^{\dagger} \circ \Lambda \circ {\mathcal{D}_{i_1}}\right)\nonumber \\
=\frac{1}{\left|\Clif\right|^{m-1}}\sum_{i_1 ... i_{m-1}} \left(\frac{1}{\left|\Clif\right|}\sum_{i_m} \delta\Lambda_{i_{m+1},m+1} \circ \left({\mathcal{D}_{i_m}}^{\dagger} \circ \Lambda \circ {\mathcal{D}_{i_{m}}} \right)\right)\circ ... \circ \left({\mathcal{D}_{i_1}}^{\dagger} \circ \Lambda \circ {\mathcal{D}_{i_1}}\right). 
\end{gather}

\end{widetext}

\noindent Since $\Clif$ is a group, if $i_1,...,i_{m-1}$ is fixed, averaging over the $i_m$ index runs through every Clifford element with equal frequency in the $\mathcal{D}_{i_{m}}$ random variable. Since $\Lambda_{i_{m+1},m+1}$ is just the error associated with the gate $\mathcal{D}_{i_{m}}^{\dagger}$, $\frac{1}{\left|\Clif\right|}\sum_{i_m} \delta\Lambda_{i_{m+1},m+1} \circ \left({\mathcal{D}_{i_m}}^{\dagger} \circ \Lambda \circ {\mathcal{D}_{i_m}} \right)$ is independent of the $i_1,...,i_{m-1}$ indices. Hence we can define

\begin{eqnarray}
\mathcal{R}_{m+1}&:=& \frac{1}{\left|\Clif\right|}\sum_{i_m} \Lambda_{i_{m+1},m+1} \circ \left({\mathcal{D}_{i_m}}^{\dagger} \circ \Lambda \circ {\mathcal{D}_{i_m}} \right) \nonumber \\
&=& \frac{1}{\left|\Clif\right|}\sum_i \Lambda_{i^{\prime},m+1} \circ \left(\mathcal{C}_i^{\dagger} \circ \Lambda \circ {\mathcal{C}_i} \right) 
\end{eqnarray}

\noindent where $\Lambda_{{i^{\prime}}, m+1}$ denotes the error that arises when the Clifford operation $\mathcal{C}_i^{\dagger}$ is applied at final time-step $m+1$. Again, using the group property of $\Clif$ we have,

\begin{equation}
\mathcal{R}_{m+1} = \frac{1}{\left|\Clif\right|}\sum_i \Lambda_{i,m+1} \circ \left(\mathcal{C}_i \circ \Lambda \circ {\mathcal{C}_i} ^{\dagger}\right). 
\end{equation}

\noindent This decoupling of $\mathcal{R}_{m+1}$ allows us to write,

\begin{widetext}

\begin{gather}
\frac{1}{\left|\Clif\right|^{m-1}}\sum_{i_1 ... i_{m-1}} \left(\frac{1}{\left|\Clif\right|}\sum_{i_m} \delta\Lambda_{i_{m+1},m+1} \circ \left({\mathcal{D}_{i_m}}^{\dagger} \circ \Lambda \circ {\mathcal{D}_{i_{m}}} \right)\right)\circ ... \circ \left({\mathcal{D}_{i_1}}^{\dagger} \circ \Lambda \circ {\mathcal{D}_{i_1}}\right) \nonumber \\
=\left(\mathcal{R}_{m+1}-\Lambda\circ \Lambda_d\right)\circ \Lambda_d^{m-1}. \label{eq:exp3}
\end{gather}

\end{widetext}

\noindent Hence combining Eq.'s (\ref{eq:Smzero}),(\ref{eq:exp1}),(\ref{eq:exp2}) and (\ref{eq:exp3}) gives,

\begin{widetext}

\begin{eqnarray}
S_m^{(0)} + S_m^{(1)} &=& \Lambda\circ \Lambda_d^m + \left(\mathcal{R}_{m+1}-\Lambda\circ \Lambda_d\right)\circ \Lambda_d^{m-1}  +  \sum_{j=2}^m \Lambda  \circ
\left( \left(\mathcal{Q}_{j} \circ \Lambda\right)_d - \Lambda_d^2  \right) \circ  \Lambda_d^{m-2} +  \Lambda \circ \Lambda_d^{m-1} \circ (\mathcal{Q}_{1}  - \Lambda_d)\nonumber \\
&=& \mathcal{R}_{m+1}\circ \Lambda_d^{m-1}  +  \sum_{j=2}^m \left(\Lambda  \circ
\left(\mathcal{Q}_{j} \circ \Lambda\right)_d \circ  \Lambda_d^{m-2}\right) + \Lambda \circ \Lambda_d^{m-1} \circ \mathcal{Q}_{1} - m\left(\Lambda \circ \Lambda_d^m\right).
\end{eqnarray}

\end{widetext}

To calculate $\Fgone := \tr\left[\left(S_m^{(0)} + S_m^{(1)}\right)( \rho_{\psi} ) E_{\psi}\right]$ we have,

\begin{gather}
\text{tr}\left(\mathcal{R}_{m+1}\circ \Lambda_d^{m-1}( \rho_{\psi} ) E_{\psi}\right)=G_{1,m+1}p^{m-1}+H_{1,m+1},
\end{gather}

\begin{gather}
\text{tr}\left(\Lambda  \circ
\left(\mathcal{Q}_{j} \circ \Lambda\right)_d \circ  \Lambda_d^{m-2}( \rho_{\psi} ) E_{\psi}\right)=A_0q_jp^{m-2}+B_0,
\end{gather}

\begin{gather}
\text{tr} \left(\Lambda \circ \Lambda_d^{m-1} \circ \mathcal{Q}_{1}( \rho_{\psi} ) E_{\psi}\right)=A_{1,1}p^{m-1}+B_0,
\end{gather}

\begin{gather}
\text{tr} \left(\Lambda \circ \Lambda_d^m( \rho_{\psi} ) E_{\psi}\right)=A_0p^m + B_0,
\end{gather}

\noindent where $G_{1,m+1}:=\text{tr}\left(\mathcal{R}_{m+1}( \rho_{\psi}-\frac{\Id}{d} ) E_{\psi}\right)$, $H_{1,m+1}:=\text{tr}\left(\mathcal{R}_{m+1}(\frac{\Id}{d})E_{\psi}\right)$, $A_{1,1}:=\text{tr}\left(\Lambda \left( \mathcal{Q}_{1}( \rho_{\psi} ) - \frac{\Id}{d}\right) E_{\psi}\right)$, $A_0$ and $B_0$ are as given in Eq.s (\ref{eq:A0def}) and (\ref{eq:B0def}), and $q_j$ is the depolarization parameter for $\left(\mathcal{Q}_{j} \circ \Lambda\right)_d$. Thus,

\begin{widetext}

\begin{gather}
\Fgone=G_{1,m+1}p^{m-1}+H_{1,m+1} + \sum_{j=2}^m(A_0q_jp^{m-2}+B_0)+A_{1,1}p^{m-1}+B_0-m\left(A_0p^m+B_0\right)\nonumber \\
=p^{m-1}\left(G_{1,m+1}+A_{1,1}-A_0p\right)+(m-1)A_0p^{m-2}\left(\frac{\sum_{j=2}^mq_j}{m-1}-p^2\right)+H_{1,m+1}. \label{eq:firstorder}
\end{gather}

\end{widetext}

Finally, we can also re-write Eq. (\ref{eq:firstorder}) as,

\begin{equation}\label{eq:model}
\Fgone =A_1(m) p^{m}  +B_1(m)+C_1  (m-1)(q(m)-p^2) p^{m-2}
\end{equation}

\noindent where,

\begin{eqnarray}
A_1(m)&=&\mathrm{Tr}\left[E_\psi \Lambda\left(\frac{\mathcal{Q}_1(\rho_\psi)}{p}-\rho_\psi+ \frac{(p-1)\openone}{pd}\right)\right]\nonumber\\
&&+\mathrm{Tr}\left[E_\psi \mathcal{R}_{m+1}\left(\frac{\rho_\psi}{p}-\frac{\openone}{pd}\right)\right]\nonumber \\
B_1(m)&=&\mathrm{Tr}\left[E_\psi \mathcal{R}_{m+1}\left(\frac{{ \openone}}{d}\right)\right]\nonumber \\
C_1&=&\mathrm{Tr}\left[E_\psi \Lambda\left(\rho_\psi-\frac{\openone}{d}\right)\right]\nonumber \\
q(m)&=&\sum_{j=2}^m q_j/(m-1),
\end{eqnarray}

\noindent and $q_j$ is the depolarizing parameter defined by

\begin{equation}
(\mathcal{Q}_{j} \circ \Lambda)_\mathrm{d}(\rho) = q_j \rho + (1 - q_j) \frac{\openone}{d}. 
\end{equation}

 \noindent We write the first order model in the form of Eq. (\ref{eq:model}) because of its similarity to that of the zeroth order model given by Eq. (\ref{eq:zerothorder}). The difference between  Eq.'s (\ref{eq:model}) and (\ref{eq:zerothorder}) is the $C_1  (m-1)(q(m)-p^2) p^{m-2}$ term contained in Eq. (\ref{eq:model}), which can be thought of as a measure of the gate-dependence of the noise.

Again, we see that the edge effects, state-preparation and measurement errors are embedded in the three coefficients $A_1(m)$, $B_1(m)$, and $C_1$. Note that the $m$ dependence in $q(m)$ and the $A_1(m)$, and $B_1(m)$ coefficients due to the last gate disappears if the errors don't change as a function of time.

%
%
%
%



\section{Neglecting Higher Orders}\label{sec:Neglectinghigherorders}

\subsection{Bounding Higher Order Perturbation Terms}

We would like to give conditions for when one is justified in stopping the expansion at some order $k$. The main idea, as expressed in Eq. (\ref{eq:Fbound}) below, is to bound the ``size" of the terms in $S_m^{(k+1)}$ and we use the $``1 \rightarrow 1$" norm on linear superoperators maximized over Hermitian inputs, denoted $\| \: \: \|_{1\rightarrow 1}^H$, to make this precise (see Sec. \ref{sec:Background}). Note that $\| \: \: \|_{1\rightarrow 1}^H$ has the following useful properties:

 \bigskip

\begin{itemize}

 \item submultiplicativity for Hermiticity-preserving superoperators,

 \item unitary invariance,

  \item $\| \Eop \|_{1\rightarrow 1}^H \leq 1$ for any quantum operation $\Eop$.

\end{itemize}

  \bigskip

\noindent Later we will discuss the motivation for using $\| \: \: \|_{1\rightarrow 1}^H$ as opposed to more familiar norms used in quantum information theory such as the diamond norm $\| \: \: \|_{\diamond}$.

From Sec. \ref{sec:norms} we have that,

\begin{gather}
\left|\Fgkpone - \Fgk\right| \nonumber \\
= \left|\tr\left[\left(\sum_{j=0}^{k+1}S_m^{(j)}\right)( \rho_{\psi} ) E_{\psi}\right] - \tr\left[\left(\sum_{j=0}^kS_m^{(j)}\right)( \rho_{\psi} ) E_{\psi}\right]\right| \nonumber \\
=  \left|\tr\left[S_m^{(k+1)}( \rho_{\psi} ) E_{\psi}\right]\right| \nonumber \\
\leq \| S_m^{(k+1)} \|_{1\rightarrow 1}^H \label{eq:Fbound}
\end{gather}

\noindent and so bounding $S_m^{(k+1)}$ provides a bound for how much the $k$ and $k+1$-order fidelities will differ. We first look at the case of stopping at first order, ie. $k=1$. There are ${m+1 \choose {2}} = \frac{(m+1)m}{2}$ second order perturbation terms in Eq. (\ref{eq:expansion}). Let us look at at a term with perturbations at $j_1$ and $j_2$ where without loss of generality we assume $j_2 > j_1$. Using the properties listed above, along with the triangle inequality, we have,

\begin{widetext}

\begin{gather}
\left\|\frac{1}{\left|\Clif\right|^m}\sum_{\vec{i_m}} \Lambda \circ \mathcal{D}_{i_m}^{\dagger}\circ  \Lambda \circ \mathcal{D}_{i_m} \circ ... \circ \mathcal{D}_{i_{j_2}}^{\dagger}\circ \delta \Lambda_{i_{j_2}}\circ \mathcal{D}_{i_{j_2}} \circ ... \circ \mathcal{D}_{i_{j_1}}^{\dagger}\circ \delta \Lambda_{i_{j_1}}\circ \mathcal{D}_{i_{j_1}} \circ ... \circ \mathcal{D}_{i_1}^{\dagger}\circ \Lambda \circ \mathcal{D}_{i_1}\right\|_{1\rightarrow 1}^H \nonumber \\
\leq \frac{1}{\left|\Clif\right|^m}\sum_{\vec{i_m}} \left\|\Lambda \right\|_{1\rightarrow 1}^H \left\| \mathcal{D}_{i_m}^{\dagger}\circ \Lambda \circ \mathcal{D}_{i_m}\right\|_{1\rightarrow 1}^H  ... \left\|\mathcal{D}_{i_{j_2}}^{\dagger}\circ \delta \Lambda_{i_{j_2}}\circ \mathcal{D}_{i_{j_2}}\right\|_{1\rightarrow 1}^H ... \left\| \mathcal{D}_{i_{j_1}}^{\dagger}\circ \delta \Lambda_{i_{j_1}}\circ \mathcal{D}_{i_{j_1}} \right\|_{1\rightarrow 1} ... \left\| \mathcal{D}_{i_1}^{\dagger}\circ \Lambda \circ \mathcal{D}_{i_1}\right\|_{1\rightarrow 1}^H \nonumber \\
=\left({\left\|\Lambda\right\|_{1\rightarrow 1}^H}\right)^{m-1}\frac{1}{\left|\Clif\right|}\sum_{i_{j_2}}\left\|\mathcal{D}_{i_{j_2}}^{\dagger}\circ \delta \Lambda_{i_{j_2}}\circ \mathcal{D}_{i_{j_2}}\right\|_{1\rightarrow 1}^H\frac{1}{\left|\Clif\right|}\sum_{i_{j_1}}\left\| \mathcal{D}_{i_{j_1}}^{\dagger}\circ \delta \Lambda_{i_{j_1}}\circ \mathcal{D}_{i_{j_1}} \right\|_{1\rightarrow 1}^H\nonumber \\
\leq \frac{1}{\left|\Clif\right|}\sum_{i_{j_2}}\left\|\mathcal{D}_{i_{j_2}}^{\dagger}\circ \delta \Lambda_{i_{j_2}}\circ \mathcal{D}_{i_{j_2}}\right\|_{1\rightarrow 1}^H\frac{1}{\left|\Clif\right|}\sum_{i_{j_1}}\left\| \mathcal{D}_{i_{j_1}}^{\dagger}\circ \delta \Lambda_{i_{j_1}}\circ \mathcal{D}_{i_{j_1}} \right\|_{1\rightarrow 1}^H \nonumber \\
=\gamma_{j_2}\gamma_{j_1} 
\end{gather}

\end{widetext}

\noindent where we define the time-dependent variation in the noise,

\begin{equation}
\gamma_j := \frac{1}{\left|\Clif\right|}\sum_i \left\|\Lambda_{i,j} - \Lambda\right\|_{1\rightarrow 1}^H.
\end{equation}

\noindent Summing over all $j_1$, $j_2$ with $j_2 > j_1$ gives,

\begin{widetext}

\begin{gather}
\left\|S_m^{(2)} \right\|_{1\rightarrow 1}^H  =  \left\|\frac{1}{\left|\Clif\right|^m}\sum_{\vec{i_m}}S_{\vec{i_m}}^{(2)} \right\|_{1\rightarrow 1}^H \nonumber \\
=\left\|\frac{1}{\left|\Clif\right|^m}\sum_{\vec{i_m}}\sum_{j_2>j_1}\Lambda \circ \mathcal{D}_{i_m}^{\dagger}\circ \delta \Lambda \circ \mathcal{D}_{i_m} \circ ... \circ \mathcal{D}_{i_{j_2}}^{\dagger}\circ \delta \Lambda_{i_{j_2}}\circ \mathcal{D}_{i_{j_2}} \circ ... \circ \mathcal{D}_{i_{j_1}}^{\dagger}\circ \delta \Lambda_{i_{j_1}}\circ \mathcal{D}_{i_{j_1}} \circ ... \circ \mathcal{D}_{i_1}^{\dagger}\circ \Lambda \circ \mathcal{D}_{i_1}\right\|_{1\rightarrow 1}^H\nonumber \\
\leq\sum_{j_2>j_1}\left\|\frac{1}{\left|\Clif\right|^m}\sum_{\vec{i_m}}\Lambda \circ \mathcal{D}_{i_m}^{\dagger}\circ \delta \Lambda \circ \mathcal{D}_{i_m} \circ ... \circ \mathcal{D}_{i_{j_2}}^{\dagger}\circ \delta \Lambda_{i_{j_2}}\circ \mathcal{D}_{i_{j_2}} \circ ... \circ \mathcal{D}_{i_{j_1}}^{\dagger}\circ \delta \Lambda_{i_{j_1}}\circ \mathcal{D}_{i_{j_1}} \circ ... \circ \mathcal{D}_{i_1}^{\dagger}\circ \Lambda \circ \mathcal{D}_{i_1}\right\|_{1\rightarrow 1}^H\nonumber \\
\leq \sum_{j_2>j_1}\gamma_{j_2}\gamma_{j_1}. \label{eq:Sbound}
\end{gather}

\end{widetext}

\noindent In terms of the fidelity we thus have from Eq.'s (\ref{eq:Fbound}) and (\ref{eq:Sbound}),

\begin{gather}
\left|\Fgtwo - \Fgone\right| \leq \sum_{j_2>j_1}\gamma_{j_2}\gamma_{j_1}.\label{eq:sufficient}
\end{gather}

\noindent Note that if the noise is time-independent then we have,

\begin{equation}
\sum_{j_2>j_1}\gamma^2 = \frac{(m+1)m}{2}\gamma^2
\end{equation}

\noindent which gives,

\begin{gather}
\left|\Fgtwo - \Fgone\right| \leq \frac{(m+1)m}{2}\gamma^2.
\end{gather}

It is straightforward to show that bounds on higher order terms go as

\begin{gather}
\left\|S_m^{(k)}\right\|_{1\rightarrow 1}^H \leq \sum_{j_k>...>j_1}\gamma_{j_k}...\gamma_{j_1}
\end{gather}

\noindent so that the difference between the $k$ and $k+1$-order fidelities is bounded by,

\begin{gather}
\left|\Fgkpone - \Fgk\right| \leq \sum_{j_k>...>j_1}\gamma_{j_k}...\gamma_{j_1}.
\end{gather}

\noindent Again if the noise is time-independent,

\begin{gather}
\left|\Fgkpone - \Fgk\right| \leq {m+1 \choose {k}}\gamma^k.
\end{gather}

We now discuss our motivation for using $\| \: \: \|_{1\rightarrow 1}^H$ as opposed to more familiar norms for distinguishing superoperators, such as the diamond norm. For any superoperator norm $\| \: \: \|$ that satisfies the properties listed above, the following inequality holds,

\begin{gather}
|\Fgkpone - \Fgk | \leq {m+1 \choose {k}} \gamma^k
 \end{gather}

\noindent where,

\begin{equation}
\gamma := \frac{1}{\left|\Clif\right|}\sum_i \left\|\Lambda_i - \Lambda\right\|
\end{equation}

\noindent and for simplicity we have assumed time-independent noise.

The above equations show that in order to give the tightest bound on the fidelity difference we would like to find the norm $\| \cdot \|$ that provides the smallest value of $\gamma$. The diamond norm $\| \cdot \|_{\diamond}$ is a candidate however by Eq. (\ref{eq:normineqs}) $\| \: \: \|_{1\rightarrow 1}^H$ is much weaker than $\| \cdot \|_{\diamond}$. Therefore $\gamma$ associated with $\| \: \: \|_{1\rightarrow 1}^H$ will be much smaller than $\gamma$ associated with $\| \cdot \|_{\diamond}$, providing a tighter bound on the fidelity difference.

\subsection{Case Where Benchmarking Fails} \label{sec:Case}

There is a simple (and highly un-physical) case for when benchmarking fails. Suppose the noise is time-independent and for each i, $\Lambda_i = \mathcal{C}_i^{\dagger}$. Then $F_g(m,\psi) = 1$ for every $m$ even though there is substantial error on each $\mathcal{C}_i$ and so benchmarking clearly fails. The key point to note here is that the noise is highly dependent on the gate chosen and so we expect that the sufficient condition derived above for ignoring higher order terms will not be satisfied (ie. $\gamma$ in this example will be far from 0). To see that this is the case, note that since $\Clif$ is a unitary 2-design it is also a unitary 1-design. Hence since $\Clif$ is $\dagger$-closed,

\begin{eqnarray}
\frac{1}{\left|\Clif\right|}\sum_{i=1}^{\left|\Clif\right|} \Lambda_i &=& \frac{1}{\left|\Clif\right|}\sum_{i=1}^{\left|\Clif\right|} \mathcal{C}_i^{\dagger}\nonumber \\
 &=& \frac{1}{\left|\Clif\right|}\sum_{i=1}^{\left|\Clif\right|} \mathcal{C}_i \nonumber \\
 &=& \Omega
\end{eqnarray}

\noindent where $\Omega$ is the totally depolarizing channel mapping every input state to the maximally mixed state $\frac{\Id}{d}$. Therefore,

\begin{equation}
\|\Lambda_i-\Lambda\|_{1\rightarrow 1}^H = \|\mathcal{C}_i^{\dagger}-\Omega\|_{1\rightarrow 1}^H.
\end{equation}

Now $\|\Lambda_i-\Lambda\|_{1\rightarrow 1}^H$ is achieved at a pure state and for any pure state $|\psi\rangle$,

\begin{equation}
(\Lambda_i-\Lambda)(|\psi\rangle\langle \psi |)=C_i^{\dagger}|\psi\rangle\langle \psi |C_i-\frac{\Id}{d}.
\end{equation}

\noindent Hence if $|\phi\rangle$ is a pure state at which $\|\Lambda_i-\Lambda\|_{1\rightarrow 1}^H$ is achieved,

\begin{eqnarray}
\|\Lambda_i-\Lambda\|_{1\rightarrow 1}^H &=& \left\|C_i^{\dagger}|\phi\rangle\langle \phi |C_i-\frac{\mathbbm{1}}{d}\right\|_1 \nonumber \\
&=& 1-\frac{1}{d}+(d-1)\frac{1}{d} \nonumber \\
&=& \frac{2(d-1)}{d}.
\end{eqnarray}

\noindent Therefore in this case,

\begin{eqnarray}
\gamma &=& \frac{1}{\left|\Clif\right|}\sum_i \left\|\Lambda_i - \Lambda\right\|_{1\rightarrow 1}^H \nonumber \\
&=& \frac{2(d-1)}{d} \geq 1
\end{eqnarray}

\noindent and so our sufficient condition is not satisfied as expected. 
%
%

It is important to note that one can devise tests for when such a pathological case is occurring. One simple test is given as follows: If the input state is $\kett$ then choose Clifford elements $\mathcal{C}_i$ that map $\kett$ to an orthogonal state in the measurement basis containing $|\psi\rangle$. For each $i$, apply $\mathcal{C}_i$ to $|\psi\rangle$ and perform the measurement. For small noise strength the output of the measurement should almost never be $\psi$, however if the noise is something close to the inverse of the gate the measurement result will be $\psi$ with high probability.

 \subsection{State Preparation and Measurement Errors}\label{sec:stateandmeas}
 
In this section we analyze the effect of state preparation and measurement errors on the benchmarking protocol. The main result is that these errors can be ignored in situations of practical relevance.
For simplicity of the discussion let us assume the gate-dependence of the noise is weak enough so that the zeroth order expression given in Eq. (\ref{eq:zerothorder}) is a valid model for the fidelity decay curve. One can obtain an estimate for $p$ as long as the fidelity curve is not constant. As state-preparation and measurement errors are accounted for in $A_0$ and $B_0$ we can obtain an estimate for $p$ regardless of the form of the state-preparation and measurement errors whenever the curve is not constant. Thus the protocol is robust against any state preparation or measurement errors unless these errors create a constant fidelity curve. It is straightforward to characterize exactly when the fidelity curve is constant. 


From Eq. (\ref{eq:zerothorder}) an exponential decay occurs if and only if $A_0$ is non-zero and $p$ lies in $(0,1)$. Hence no decay occurs if and only if one of $p = 0$, $p = 1$ or $A_0 = 0$ occurs. We look at each case separately.

\medskip

$\underline{p=0}$: This occurs if and only if $\Lambda$ is the totally depolarizing channel and in this case the fidelity is constant at $B_0 = \frac{\tr (E_{\psi})}{d} \leq \frac{1}{d}$. Since we have assumed small gate-dependence, this case is only possible if most of the errors are approximately centred around the totally depolarizing channel with little variation. This situation is of little practical relevance since the gate operations being characterized are usually reasonably precise.

\medskip

$\underline{p=1}$: This case corresponds to $\Lambda$ being the identity channel which means all gates are perfect. Again, in practice this situation is unlikely as the implementation of any gate will have some associated error. Note that in this case the fidelity is equal to $A_0 + B_0$ which is just $\tr(\Lambda(\rho_{\psi})E_{\psi}))=\tr(\rho_{\psi}E_{\psi})$. Hence 
the constant decay curve is a measure of the overlap between the imperfect input state and imperfect POVM element.

\medskip

$\underline{A_0=0}$: The case $A_0=0$ occurs if and only if 

\begin{equation}
\tr(E_{\psi}\Lambda(\rho_{\psi}))=\tr \left(E_{\psi}\Lambda\left(\frac{\openone}{d}\right)\right).\label{eq:Azeroeqzero}
\end{equation}

\noindent Thus $\Lambda(\rho_{\psi}))$ and $\Lambda\left(\frac{\openone}{d}\right)$ have the same probability of producing the output ``$\psi$" from the measurement. Since gates are reasonably precise in practice, this situation occurs when at least one of state preparation or measurement has substantial error. Note that the fidelity will be equal to $B_0$ in this case and so can take any value in $[0,1]$. 

From the above three cases, the only one that depends upon state preparation or measurement errors is the case $A_0=0$. Since this case occurs when at least one of state preparation or measurement errors has substantial error it is unlikely to arise in practice. This discussion shows that a constant fidelity decay curve can only occur in extreme cases and so it is safe to assume the protocol is independent of state preparation and measurement errors.

 \section{Average Error Rate and the Diamond Norm}\label{sec:Normrelationships}

In terms of connections between the average error rate $r$ and relevant fault-tolerant measures of error, it is natural to ask how the error rate $r$ between $\Lambda$ and $\mathcal{I}$ is related to the diamond norm between $\Lambda$ and $\mathcal{I}$. In general an explicit relationship will be impossible to obtain, however we show that in certain cases that are relevant in various fault-tolerant noise models we can obtain such a relationship. First we give a new proof of a previously established result~\cite{Sac05} for calculating the diamond norm distance between generalized Pauli channels. The proof we present here illustrates how one can apply a semidefinite program to calculate the diamond norm distance between quantum channels~\cite{Wat09}. Ideally, this proof technique could be used to either explicitly calculate or place bounds on the diamond norm distance between more general classes of quantum channels. This could allow for obtaining further relationships between $r$ and the diamond norm distance which hold in more general cases.

 \subsection{Calculating the Diamond Norm Distance Between Generalized Pauli Channels}

Suppose $\Eop_1$ and $\Eop_2$ are Pauli channels, or more generally any channels with Kraus operators given by an orthogonal (normalized to d) basis of unitary operators $\{P_i\}_{i=0}^{d^2}$ (which we call generalized Pauli channels),

\begin{gather}
\Eop_1 (\rho) = \sum_{i=0}^{d^2-1} q_i P_i \rho P_i^{\dagger}
\end{gather}

\begin{gather}
\Eop_2 (\rho) = \sum_{i=0}^{d^2-1} r_i P_i \rho P_i^{\dagger}.
\end{gather}

\noindent Define the vector $\vec{v}$ of length $d^2$ by

\begin{gather}
v_i=q_i-r_i
\end{gather}

\noindent for all i $\in \{0,...,d^2-1\}$. Then,

\begin{gather}\label{eq:diam}
\|\Eop_1 - \Eop_2\|_{\diamond} = \|\vec{v}\|_1 =\sum_{i=0}^{d^2-1}|v_i|.
\end{gather}


To prove Eq. (\ref{eq:diam}) using the semidefinite program in~\cite{Wat09} first note that $\Phi=\Eop_1 - \Eop_2$ has action,

\begin{gather}
\Phi(\rho) = \sum_{i=0}^{d^2-1} (q_i-r_i) P_i \rho P_i^{\dagger}.
\end{gather}

\noindent The semidefinite program has the following primal and dual problems:

\medskip

Primal problem: Maximize $\langle J(\Phi),W\rangle$ subject to $W \leq \Id_d \otimes \rho$, $W \in \text{Pos}\left(L\left(\mathbb{C}^d \otimes \mathbb{C}^d\right)\right)$, $\rho \in D\left(L\left(\mathbb{C}^d\right)\right)$,

\medskip

Dual problem: Minimize $\|\tr_1(Z)\|_{\infty}$ subject to $Z \geq J(\Phi)$, $Z \in \text{Pos}(L\left(\mathbb{C}^d \otimes \mathbb{C}^d\right))$,

\medskip

\noindent where $J(\Phi)$ is the Choi matrix~\cite{Choi} of $\Phi$. If $\alpha$ and $\beta$ are the solutions to the primal and dual problems then the case that $\alpha = \beta$ is called \textit{strong duality}. It is shown in~\cite{Wat09} that the above semidefinite program always has the property of strong duality and the solution to the program is $\alpha = \frac{1}{2}\|\Eop_1-\Eop_2\|_{\diamond}$. Note also that it is always the case that $\alpha \leq \beta$.

By definition,

\begin{gather}
J(\Phi)= d \Phi \otimes \mathcal{I}(|\psi_0\rangle\langle \psi_0 |) \nonumber \\
= d \sum_{i=0}^{d^2-1} (q_i-r_i) P_i \otimes \Id |\psi_0\rangle\langle \psi_0 | P_i^{\dagger} \otimes \Id.
\end{gather}

\noindent Noting that $\{|\psi_i\rangle := P_i \otimes \Id |\psi_0\rangle\}_{i=0}^{d^2-1}$ forms an orthonormal basis of maximally entangled states for $\mathbb{C}^d \otimes \mathbb{C}^d$, which we call the generalized Bell basis (GBB), we have that $J(\Phi)$ is diagonal when written in GBB with diagonal elements (eigenvalues) $d(q_i-r_i)$. Let $\Pi_+$ denote the projector onto the eigenspace with non-negative eigenvalues and $\Pi_-$ denote the projector onto the eigenspace with negative eigenvalues.

For the primal problem let $W=\frac{\Pi_+}{d}$ and $\rho = \frac{\Id}{d}$. Then

\begin{gather}
\langle J(\Phi),W\rangle = \sum_{k: q_k-r_k \geq 0}q_k-r_k = \frac{1}{2}\sum_k |q_k-r_k| = \frac{1}{2}\|\vec{v}\|_1.
\end{gather}

\noindent Thus $\alpha \geq \frac{1}{2}\|\vec{v}\|_1$.

For the dual problem take $Z=d \Pi_+J(\Phi)\Pi_+$ which is just $\sum_{k: q_k-r_k \geq 0}(q_k-r_k) |\psi_k\rangle \langle \psi_k|$ and note $Z \geq J(\Phi)$. Moreover, $\tr_1(Z) = d\left(\sum_{k: q_k-r_k \geq 0}q_k-r_k\right)\frac{\Id}{d}$ and so

\begin{gather}
\|\tr_1(Z)\|_{\infty} = \left(\sum_{k: q_k-r_k \geq 0}q_k-r_k\right) = \frac{1}{2}\|\vec{v}\|_1.
\end{gather}

\noindent Thus $\alpha \leq \frac{1}{2}\|\vec{v}\|_1$ which implies $\alpha = \frac{1}{2}\|\vec{v}\|_1$ and $\|\Eop_1-\Eop_2\|_{\diamond} = \|\vec{v}\|_1$ as desired.

As a simple corollary to Eq. (\ref{eq:diam}) note that if $\Eop_1$ and $\Eop_2$ are depolarizing channels with fidelity parameters $p_1$ and $p_2$ respectively then,

\begin{gather}
\|\Eop_1-\Eop_2\|_{\diamond}=\frac{2|p_1-p_2|(d^2-1)}{d^2}.
\end{gather}

\noindent To see this note that

\begin{gather}
q_0 = \frac{(d+1)\overline{F_{\Eop_1,\mathcal{I}}}-1}{d} = \frac{(d+1)\left(p_1 + \frac{1-p_1}{d}\right)-1}{d}\nonumber \\
=\frac{(d^2-1)p_1 + 1}{d^2}
\end{gather}

\noindent and similarly,

\begin{gather}
r_0 =\frac{(d^2-1)p_2 + 1}{d^2}.
\end{gather}

\noindent Thus for every $1\leq i \leq d^2-1$,

\begin{gather}
q_i = \frac{1-q_0}{d^2-1} = \frac{1-p_1}{d^2}
\end{gather}

\noindent and

\begin{gather}
r_i = \frac{1-r_0}{d^2-1} = \frac{1-p_2}{d^2}.
\end{gather}

\noindent So,

\begin{eqnarray}
\|\Eop_1-\Eop_2\|_{\diamond} &=& \|v\|_1 \nonumber \\
&=& |q_0-r_0| + \sum_{i=1}^{d^2-1}|q_i-r_i|\nonumber \\
&=& \left|\frac{(d^2-1)p_1 + 1}{d^2} - \left(\frac{(d^2-1)p_2+1}{d^2}\right)\right| \nonumber \\
& \: \: & + (d^2-1)\left|\frac{1-p_1}{d^2}-\left(\frac{1-p_2}{d^2}\right)\right|\nonumber \\
&=& 2 \frac{(d^2-1)|p_1-p_2|}{d^2}.
\end{eqnarray}

 \subsection{Relating the Diamond Norm and Error Rate in Benchmarking}

Now suppose that $\Eop_2 = \mathcal{I}$ in Eq. (\ref{eq:diam}). Then, $r_0=1$ and for every $1 \leq i \leq d^2-1$, $r_i=0$. Hence in this case,

\begin{gather}
\|\Eop_1-\mathcal{I}\|_{\diamond}=\|\vec{v}\|_1 = |q_0-1| + 1-q_0 = 2(1-q_0).
\end{gather}

\noindent We know that $q_0$ is related to the average fidelity of $\Eop_1$, $\overline{F_{\Eop_1,\mathcal{I}}}$, by

\begin{gather}
\overline{F_{\Eop_1,\mathcal{I}}} = \frac{q_0d+1}{d+1}
\end{gather}

\noindent and so,

\begin{gather}
\| \Eop_1 - \mathcal{I}\|_{\diamond} = \frac{2(d+1)(1 - \overline{F_{\Eop_1,\mathcal{I}}})}{d}.
\end{gather}

\noindent Therefore in the case of randomized benchmarking (where we define the error rate $r = 1 - \overline{F_{\Lambda,\mathcal{I}}}$) if $\Lambda$ is a generalized Pauli channel, $r$ and $\| \Lambda - \mathcal{I}\|_{\diamond}$ are related by,

\begin{gather}
\| \Lambda - \mathcal{I}\|_{\diamond} = 2\frac{(d+1)r}{d}.
\end{gather}

 \section{Scalability of the Protocol}\label{sec:Scalability}

In this section we fill in the details of the scalability proof of our RB protocol that was briefly outlined in~\cite{MGE}. First, we note that the size of the Clifford group scales as $2^{O\left(n^2\right)}$ and so the number of sequences of length m scales as $2^{mO\left(n^2\right)}$. Hence if full averaging over the Clifford group is required for each sequence length, our protocol does not scale well in either of $n$ or $m$. As mentioned in~\cite{MGE}, there are three obstacles to overcome in order for the above protocol to be scalable:

\bigskip

\noindent 1. Sequence length: Since the number of sequences of length $m$ scales as $2^{mO\left(n^2\right)}$, averaging over \emph{all} sequences for each $m$ is clearly inefficient.

\medskip

\noindent 2. Uniform sampling: Since the size of the Clifford group scales as $2^{O\left(n^2\right)}$, sampling directly from a list of all Clifford elements becomes impossible for large $n$ (writing down every element is inefficient in $n$).

\medskip

\noindent 3. Implementing Clifford operations: In practice, one can only implement a generating set for the Clifford group. Hence even if random sampling can be accomplished there must be a scalable method for implementing each Clifford using only this generating set.

\bigskip

\noindent We now describe how to overcome each of the above obstacles.



\bigskip

\noindent \underline{Solution to 1}: From Eq. (\ref{eq:avgdepseq}), $\Fg$ is the uniform average of the random variable


\begin{gather}
\Fgvec := \tr\left(S_{\vec{i_m}} ( \rho_{\psi} ) E_{\psi}\right) \nonumber \\
= \tr\Big(\Lambda_{i_{m+1},m+1} \circ \mathcal{C}_{i_{m+1}} \circ   ... \circ \Lambda_{i_1,1}  \circ \mathcal{C}_{i_1} ( \rho_{\psi} ) E_{\psi}\Big)
\end{gather}

\noindent over $\left|\Clif\right|^m$ sequences $(i_1,...,i_m)$. The benchmarking protocol requires choosing a sequence at random, evaluating the above fidelity, repeating for many sequences, and taking the average of the results.

Let $S_k(m,\kett)=\frac{\Fgvec + .... + \Fgvec}{k}$ be the normalized k-fold sum of the random variable $\Fgvec$ and note that $\mathbbm{E}[S_k(m,\kett)]=\Fg$. A probablistic bound on $\left|S_k(m,\kett)-\Fg\right|$ is given by H\"oeffding's inequality,

\begin{eqnarray}
\mathbbm{P}\left(\left|S_k(m,\kett)-\Fg\right| \geq \epsilon \right) &\leq& 2 e^{\frac{-2(k\epsilon)^2}{k\left(b-a\right)^2}} \nonumber \\
&=& 2 e^{\frac{-2k\epsilon^2}{\left(b-a\right)^2}}
\end{eqnarray}

\noindent where $[a,b]$ is the range of $\Fgvec$. Since $\Fgvec$ is a fidelity it must lie in $[0,1]$ (in reality it will lie in a much smaller interval, for now we continue to assume it lies in $[a,b]\subseteq [0,1]$). Suppose we want

\begin{equation}
\mathbbm{P}\left(\left|S_k(m,\kett)-\Fg\right| \geq \epsilon \right)\leq \delta
\end{equation}

\noindent where $\epsilon$ represents the accuracy of the estimate and $1-\delta$ represents the desired confidence level. We can find how many trials one needs to perform to obtain this accuracy by setting $\delta=2 e^{\frac{-2k\epsilon^2}{\left(b-a\right)^2}}$ and solving for k,

\begin{gather} \label{eq:notrials}
k=\frac{\ln \left(\frac{2}{\delta}\right)(b-a)^2}{2\epsilon ^2}.
\end{gather}

\noindent Note that k is explicitly independent of m and n which provides a solution to 1.

 It is instructive to obtain an estimate of the size of $k$ for realistic parameter values of $\delta$ and $\epsilon$. Since $1-\delta$ represents our desired confidence level we set $\delta = 0.05$. Fault-tolerance provides a wide range for the error tolerance of a physical (0-level) gate in the fault-tolerant construction. The value of the error tolerance depends on both the coding scheme as well as the noise model and typical values lie somewhere between $10^{-6}$ and $10^{-2}$. Let us assume that the physical gates have errors on the order of $10^{-4}$. Intuitively, since the fidelity curve decays in sequence length it is reasonable to assume that $\epsilon$ can be relaxed as $m$ grows large. Similarly, $b-a$ can be assumed to be relatively small for small values of $m$ but will converge to $1-\frac{1}{d}$ as $m$ grows large. As a result both $b-a$ and $\epsilon$ have an \emph{implicit} dependence on $m$ and this implicit dependence is advantageous when choosing $\epsilon$ for large values of $m$. Let us assume $m=100$ and a fidelity decay curve that is well-approximated by an exponential. Then we expect fidelity values on the order of $0.99$ at this value of $m$ and so we take $\epsilon=10^{-3}$, $b-a=0.2$. With these values for $\epsilon$, $\delta$ and $b-a$ we get,

\begin{eqnarray}
k&=&\frac{\ln \left(\frac{2}{0.05}\right)(0.2)^2}{2(10^{-3}) ^2} \nonumber \\
&\sim& 7 \times 10^4.
\end{eqnarray}

\noindent While this number is large it is independent of $n$ and thus compares favourably with quantum process tomography which scales as $16^n$. As a direct comparison, performing process tomography on a 4 qubit system already requires $65 536$ measurements.

\bigskip

\noindent \underline{Solution to 2}:

For the second problem we present a method to scalably sample uniformly from the full Clifford group that utilizes the symplectic representation of the Clifford group (see Ref's~\cite{Dan05,DDM}). Since the Clifford group is the normalizer of the Pauli group, every Clifford element is completely determined by its action under conjugation on the Pauli group. In particular, since the Pauli group is generated by the set of all $X_i$ and $Z_i$ (the label $i$ refers to $X$ or $Z$ being in the $i$'th position with identity operators elsewhere), an element of the Clifford group is completely determined by its action on this set. In the symplectic representation this corresponds to each Clifford element $\mathcal{Q}$ being associated uniquely to a $2n$ by $2n$ binary symplectic matrix $C$ and length $2n$ binary vector $h$ which records negative signs in the images of $X_i$ and $Z_i$. The only constraints on $\mathcal{Q}$ are that commutation relations and Hermiticity of the generating set must be preserved under $\mathcal{Q}$. Hence we can construct a random Clifford element $\mathcal{Q}$ by inductively constructing a random symplectic matrix $C$ and vector $h$. 

Since $h$ corresponds to keeping track of negative signs, the binary entries of $h$ can be chosen uniformly at random. $C$ is inductively constructed column by column where the first $n$ columns correspond to the images of $X_1$ through $X_n$, and the last $n$ columns correspond to the images of $Z_1$ through $Z_n$ (all of which are written in binary notation as in~\cite{DDM}). Preservation of commutation relations is phrased through the symplectic inner product and so at each step one chooses the new column by finding a random solution to a system of linear equations which represents the inner product conditions. Since randomly choosing $2n$ elements of the Pauli group that satisfy the required commutation relations is equivalent to inductively choosing random solutions to $2n$ sets of linear equations (which requires $O\left(n^3\right)$ operations), we can produce a random Clifford element in $O\left(n^4\right)$ (classical) operations.

\bigskip

\noindent \underline{Solution to 3}: Any Clifford element can be decomposed into a sequence of $O\left(n^2\right)$ one and two-qubit generators in $O\left(n^2\right)$ time~\cite{DDM} (alternatively, there are slower methods which produce a ``canonical" decomposition into $O\left(n^2/\log n\right)$ generators~\cite{AG}). We describe this method which again utilizes the symplectic representation of the Clifford group. As mentioned above, every Clifford element $\mathcal{Q}$ is represented up to phase by a binary, symplectic matrix $C$ and a binary vector $h$. The main goal is to decompose $C$ into generators as the negative signs represented by $h$ can be accounted for via multiplication by single-qubit Pauli operators. The main theorem used in the decomposition of Clifford elements is theorem 4 of~\cite{DDM} which states that if $C$ is a binary symplectic matrix then $C$ can be decomposed as a product of five binary symplectic matrices, which we denote by $T_1$ through $T_5$. 



These symplectic matrices can be decomposed into symplectic matrices representing $1$ and $2$-qubit Clifford operations that correspond to Hadamard's, single qubit $\frac{\pi}{2}$-rotations about $\sigma_Z$, two-qubit $\frac{\pi}{2}$-rotations about $\sigma_Z \otimes \sigma_Z$, two-qubit permutation operations and CNOT operations. The overall discussion can be condensed into the following main result:

\medskip

\noindent Main Result: Every Clifford operation $\mathcal{Q}$ can be realized by a sequence of one and two-qubit Clifford operations which consists of the following six rounds of operations:

\bigskip

1. An initial round of single-qubit Pauli operators,

\medskip

2. Applying a sequence of CNOT and two-qubit permutation operations,

\medskip

3. Applying a sequence of $\frac{\pi}{2}$ rotations about $\sigma_Z \otimes \sigma_Z$ followed by a sequence of $\frac{\pi}{2}$ rotations about $\sigma_Z$,

\medskip

4. Applying Hadamard operations,

\medskip

5. Applying a sequence of $\frac{\pi}{2}$ rotations about $\sigma_Z \otimes \sigma_Z$ followed by a sequence of $\frac{\pi}{2}$ rotations about $\sigma_Z$,

\medskip

6. Applying a final round of CNOT and two-qubit permutation operations.

\bigskip

\noindent Note that the operations within each of the rounds 3, 4 and 5 all commute and can be performed in any order.

The time-complexity in decomposing a symplectic matrix into the sequence of one and two-qubit Clifford operations given above is $O(n^3)$ since one needs to solve linear systems of equations to obtain $T_1$ through $T_5$. In many cases one would like to have a decomposition of a Clifford element into a particular generating set for the Clifford group, such as $G_n:=\{$H,S,CNOT$\}$ which consists of Hadamard's (H) and phase gates (S) on each qubit, as well as CNOT gates on all pairs of qubits. There are $n^2 + n$ elements in $G_n$ and it is a straightforward process to decompose the operations in 1 through 6 above into H, S and CNOT gates.

In total, for an n-qubit system, we can efficiently choose Clifford gates uniformly at random and decompose each gate into a canonical subsequence of elements from the generating set $G_n$. The total time complexity of these two procedures is $O\left(n^4\right) + O\left(n^3\right) = O\left(n^4\right)$. The number of trials $k$ one needs to perform to estimate $\Fg$ to an accuracy $\epsilon$ with probability at least $1-\delta$ is  given by Eq. (\ref{eq:notrials}) which is independent of m and n. Thus if we perform the protocol for $R$ different values of m, the total time complexity is

\begin{gather}\label{eq:timecomplexity}
\frac{O\left(n^4\right)\cdot R\ln(2/\delta)}{2\epsilon^2} 
\end{gather}

\noindent which implies the protocol is scalable in $n$.

\section{Discussion}\label{sec:Discussion}


We have shown that randomized benchmarking provides a scalable method for benchmarking the set of Clifford gates. The protocol allows for time and gate-dependent noise and the fitting models for the fidelity function take into account state preparation and measurement errors. In addition to providing an estimate of the average fidelity across all Clifford gates, the first order model provides a measure of the gate-dependence of the noise. 

We have provided here rigorous proofs of both the conditions for the validity of the protocol, as well as the scalability of the protocol in the number of qubits $n$ comprising the system. We have also established an exact relationship between the average fidelity estimate provided by the protocol and a stronger characterization of the average error operator strength given by the diamond norm for the case of random Pauli errors. The proof of this relationship utilizes a semidefinite program for computing the diamond norm~\cite{Wat09} which has the potential to establish further connections between these two notions of error strength.

While benchmarking the full unitary group would be ideal, this is a provably inefficient task since just generating a Haar-random unitary operator is inefficient in $n$. On the other hand as we have shown here benchmarking the Clifford group is an efficient task. It is not difficult to see that benchmarking the Clifford group provides significant information for both fault-tolerant quantum computation as well as obtaining a benchmark for a generating set of the full unitary group. First, any realistic implementation of a quantum computer will have to take advantage of error-correction codes in order to perform fault-tolerant quantum computation. The fact that most of the codes used in fault-tolerant theory are stabilizer codes implies that the encoding and decoding operations that have to be performed can be chosen to be Clifford operations. Hence a benchmark of Clifford operations provides direct information regarding the robustness of these encoding/decoding schemes. 

Second, the unitary group can be generated by adding just one single-qubit rotation not in the Clifford group (for instance the $\frac{\pi}{8}$-gate). Hence a benchmark for the Clifford group can actually provide useful information regarding a benchmark for a generating set of the full unitary group.
In addition, it has been shown that any unitary operation can be implemented using Clifford gates, a single-qubit ancilla state called a magic state~\cite{BK} and measurements in the computational basis. Hence in this model of quantum computation the only gates that need to be benchmarked for universal quantum computation are Clifford gates.

Various interesting questions and comments arise from the benchmarking analysis presented here. First, there is a key point to emphasize regarding the zeroth and first order fitting models. As depicted in~\cite{MGE} there exist physically relevant noise models for which when the true value of the depolarization fidelity parameter $p$ is used, the first order model fits the experimental data much better than the zeroth order model. However, it may be the case that a least squares fitting procedure using the functional form of the zeroth order model produces a very good fit to the experimental data, albeit producing an incorrect value for $p$. Therefore in order to obtain a more accurate value for $p$ one should always use the first order fitting model unless prior knowledge of the noise indicates that it is effectively gate-independent.

It will be useful to obtain a better understanding for when a least squares fitting procedure using the zeroth order model produces a value for $p$ that is close to its true value. Clearly in the gate-independent case the zeroth order model fits the fidelity decay curve exactly. Moreover for weakly gate-dependent noise one can see from our continuity argument that the zeroth order model is still a sufficient fitting function for the fidelity decay curve. Hence the most interesting case to analyze is when there is a non-negligible amount of gate-dependence in the noise and the condition for using the first order model to fit the decay curve is satisfied.
A useful test that would indicate gate-dependence in the noise, and thus the validity of the value of $p$ obtained from fitting to the zeroth order model, is to perform the least squares fitting procedure using both the zeroth and first order fitting models. If the estimates of $p$ obtained in each case differ significantly then the zeroth order model must be a poor choice of fitting function even though it may fit the data well. In this case the noise must have a strong gate-dependence because otherwise $q-p^2$ would be small which implies the two fitting functions would produce similar estimates for $p$.

An interesting question is how to extract a meaningful average error rate over a generating set of the Clifford group, for instance $G_n$ defined previously, from the average error rate $r$ over the entire Clifford group. One might argue that benchmarking a generating set for the Clifford group is sufficient for benchmarking the full Clifford group, however it is entirely plausible that noise correlations between the $n$ physical qubits creates large errors on elements of $\Clif$, even when the errors on the generating set can be controlled~\cite{MSB}. In fact an assumption that is often made in fault-tolerant estimates is that the correlation in noise between qubits is either small or can be ignored.
%

With regards to scalability, while we have shown the protocol itself is scalable in $n$, a useful direction for further research would be an analysis of how the sufficient condition of weak average variation of the noise depends on $n$. As previously noted, the noise associated to a multi-qubit Clifford element is given by the noise associated to the sequence of generators comprising the Clifford. A determination of whether these noise operators continue to satisfy the sufficient condition when it is met for small numbers of qubits will be useful for understanding the applicability of the protocol.

Rigorous fault-tolerant analyses sometimes invoke the diamond norm as a measure of the error strength rather than the weaker characterization provided by the average fidelity. Hence it is desirable to find relationships between these two quantities that is more general than the special case of random Pauli errors presented here. As mentioned above, the semidefinite program we have used to deduce the relationship appears to be a promising tool for further research in this area. From the expression given in Eq. (\ref{eq:Diamond2}) one can see that the diamond norm is essentially a ``worst-case" maximization over input (entangled) states. In quantum computation it is the case that the measure of accessible states (states that can be reached in polynomial time using a generating set for the unitary group) is equal to 0. Hence there is a high probability that the maximization criteria demanded by the diamond norm is a much stronger condition than necessary for understanding the strength of the errors affecting the computation. This point becomes even more relevant for an algorithm-specific (ie. non-universal) quantum computer. An interesting direction of further research is to provide precise conditions for when the average fidelity provides an indication or bound on the error strength in terms of stronger characterizations such as the diamond norm. 

Additionally, if one were able to obtain an estimate of the minimum gate fidelity from knowledge of the average fidelity they could use the direct relationship between the minimum gate fidelity and diamond norm given by Eq. (\ref{eq:minfidtodiamond}) to obtain information about the error strength in terms of the diamond norm. A result that may be useful in this direction of research is the ``concentration of measure effect" of the gate fidelity which implies that 
as $n$ increases, the measure of the set of states which produce a fidelity close to the minimum yet far from the average is exponentially small in $n$~\cite{MBKE, MAG11}.

\end{document}